\documentclass[review]{elsarticle}

\usepackage{lineno,url}
\modulolinenumbers[5]

\usepackage{amsmath,amssymb}
\usepackage{bbm}
\usepackage{graphicx}
\usepackage{algorithmic}
\usepackage{multirow}
\usepackage{color}
\usepackage{textcomp}

\newtheorem{defi}{Definition} 
\newtheorem{theo}{Theorem}  
\newtheorem{prop}{Proposition}
\newtheorem{cor}{Corollary}
\newtheorem{ass}{Modeling Assumption}
\newtheorem{lem}{Lemma}

\newcommand\eatpunct[1]{}

\journal{Medical Image Analysis}

\biboptions{authoryear,square}

\begin{document}

\begin{frontmatter}

\title{Hierarchical segmentation using equivalence test (HiSET): Application to DCE image sequences}


\author[paris5,intrasense]{Fuchen Liu}

\author[paris5,hegp,s970]{Charles-Andr\'e Cuenod}

\author[paris6,tenon,s970]{Isabelle Thomassin-Naggara}

\author[intrasense]{St\'ephane Chemouny}

\author[paris5,biostm]{Yves Rozenholc\corref{mycorrespondingauthor}}
\cortext[mycorrespondingauthor]{Corresponding author}
\ead{yves.rozenholc@parisdescartes.fr}

\address[paris5]{Universit\'e Paris Descartes - USPC, France}
\address[biostm]{Faculté de Pharmacie de Paris - EA BioSTM, France}
\address[paris6]{Universit\'e Pierre et Marie Curie - Sorbonne Université, France}
\address[hegp]{H\^opital Europ\'een Georges Pompidou (HEGP) - APHP, France}
\address[s970]{UMR-S970, PARCC, France}
\address[tenon]{H\^opital Tenon - APHP, France}
\address[intrasense]{Intrasense\textsuperscript{{\rm \tiny\textregistered}}, France}

\begin{abstract}
{\it Dynamical contrast enhanced} (DCE) imaging allows non invasive access to tissue micro-vascularization. 
It appears as a promising tool to build imaging biomarkers for diagnostic, prognosis or anti-angiogenesis treatment monitoring of cancer.
However, quantitative analysis of DCE image sequences suffers from low {\it signal to noise ratio} (SNR). 
SNR may be improved by averaging functional information in a large region of interest when it is functionally homogeneous.

We propose a novel method for automatic segmentation of DCE image sequences into functionally homogeneous regions, called {\it DCE-HiSET}.
Using an observation model which depends on one parameter $a$ and is justified {\it a posteriori}, DCE-HiSET is a hierarchical clustering algorithm. It uses the $p$-value of a multiple equivalence test as dissimilarity measure and consists of two steps. The first exploits the spatial neighborhood structure to reduce complexity and takes advantage of the regularity of anatomical features, while the second recovers (spatially) disconnected homogeneous structures at a larger (global) scale.

Given a minimal expected homogeneity discrepancy for the multiple equivalence test, both steps stop automatically by controlling the Type I error. This provides an adaptive choice for the number of clusters. Assuming that the DCE image sequence is functionally piecewise constant with signals on each piece sufficiently separated, we prove that DCE-HiSET will retrieve the exact partition with high probability as soon as the number of images in the sequence is large enough. The minimal expected homogeneity discrepancy appears as the tuning parameter controlling the size of the segmentation. DCE-HiSET has been implemented in C++ for 2D and 3D image sequences with competitive speed.  

\end{abstract}

\begin{keyword}
DCE imaging \sep automatic clustering \sep hierarchical segmentation \sep equivalence test 
\end{keyword}

\end{frontmatter}


\section{Introduction}

\subsection{DCE imaging}
DCE (dynamic contrast enhanced) imaging using computed tomography (CT), magnetic resonance imaging (MRI) or ultrasound imaging (US) appears promising as it can monitor the local changes in microcirculation secondary to the development of new vessels (neo-angiogenesis). DCE-MRI and DCE-CT (called also CT-perfusion) have been extensively tested alone or in combination with other techniques \citep{cuenod1} in pathological conditions such as cancer, ischemia and inflammation, in various tissues including brain \citep{cuenod2}, breast \citep{cuenod3}, prostate \citep{cuenod4}, heart \citep{cuenod5,cuenod6}, kidney \citep{cuenod7}, liver \citep{cuenod8,cuenod9}, genital organs, gastrointestinal tract, bone \citep{cuenod10} and placenta \citep{cuenod11}. They show a great potential to: 1/ detect and characterize lesions \citep{cuenod4,cuenod12,cuenod13}; 2/ personalize treatment including new targeted drugs, radiotherapy and mini invasive surgery; 3/ monitor and optimize treatments during the follow-up \citep{cuenod14,cuenod15} or after heart, liver or kidney transplants \citep{cuenod16}. 

DCE imaging follows the local distribution of a contrast agent after intravenous injection, using sequential acquisition \citep{Ingrisch:2013aa}. Taken at the same cross-section of the patient during the entire acquisition, the images of the {\it DCE image sequence} discretely reveal, at each voxel, a {\it time curve} (TC) made of noisy intensities. TCs analysis involves either hardly reproducible abstract parameters \citep{PAAUC2004,PAME2005,PATTP2004} or physiological parameters after a deconvolution with respect to the patient's Arterial Input Function (AIF)\footnote{The AIF is the main blood flow arriving to the tissue of interest, often the aorta or a large artery} \citep{PABrix2004, Chung:2016aa, PATofts1997, PATofts1995,Axel:1980aa,fieselmann2011, Ostergaard:1996aa,PABayes2006,DCEMCMC2009,LaplaceDeconv2014}.

\subsection{Limitations}\label{sec:issues}
However, DCE imaging suffers from two technical issues, that hamper its use: movements and low SNR.
Movements (such as breathing, cardiac motion and non-periodic motions like bowel peristalsis) can be addressed by motion correction \citep{glocker-ar,sotiras-miccai}. Movements and registration is not our concern and we assume further that the DCE image sequences have already been registered. 
Low SNR, which is our concern, affects any estimation. 
It is a consequence of minimizing the total X-ray dose received by the patient during this multi-image acquisition in DCE-CT or of the short delay between two consecutive images in DCE-MRI. 


To improve SNR, either TCs are averaged over large regions of interest (ROI) manually drawn by the radiologist or denoised with spatial filtering techniques  (e.g. PCA). 
However, large ROI could result in a lack of homogeneity. In this case, the gain in variance resulting from averaging is obtained at the price of a high bias which may compromise further analysis. Similarly, filtering techniques applied to the observed noisy TCs could trigger a tricky trade-off between information and noise when the noise structure has heavy tails or when the TC shows high frequencies. The latter is a typical consequence of injecting the contrast agent as a bolus. From a statistical point-of-view, as a result of the bias-variance decomposition of the estimation risk, this issue can be solved using adaptive statistical procedures (e.g. threshold wavelet decomposition or multi-scale comparison) with respect to the (unknown) smoothness of the (unknown) underlying signal. However, as the functional information in DCE sequences results from a convolution with the AIF, the true signals can be expected to be smooth enough. In this context, a multi-scale comparison, known to be adaptive with respect to H\"older regularity, provides a good bias-variance trade-off while having low level of complexity.\medskip

The fixed cross section of a DCE image sequence defines a fixed spatial (2D or 3D) domain made up of voxels. At each voxel, a TC is observed in the time domain, leading to the 2D+T or 3D+T representations. After registration, a DCE image sequence reveals tissues (organs, vessels, bones, muscles, fat, etc) as having homogenous functional properties. Hence, tissues can be considered as spatially static functional objects. As shown by \citep{zheng2009step} for breast tumors and on our real data in Section \ref{sec:exp}, even heterogeneous tumors or metastases have only a small number of functional behaviors. Thus, their homogeneous sub-parts can also be treated as static functional objects. It leads to segmenting the DCE image sequence into regions (of voxels) showing homogeneous TCs. Within each of these regions, SNR can be improved by averaging without the loss of temporal information. This alternative is known as {\it DCE image sequence segmentation}. %

Segmentation is known to be a useful tool for common image analysis to get SNR improvement, it is an even more important preprocessing steps for DCE-imaging analysis. Indeed, in addition to the expected SNR improvement, it allows to: 1/ reveal the functional anatomy that is hardly visible on static images \citep{cuenod18}, therefore allowing the correct selection of different tissues \citep{cuenod20,cuenod21}; 
2/ summarize the functional information, hence reducing the amount of parameter extractions; 
3/ increase the conspicuity of the images and facilitate the fit of the TCs \citep{cuenod17,cuenod19,cuenod22}; 
4/ analyze genuine tissues and obtain correct parameters;
5/ provide and guarantee the detection, measurement and characterization of lesions  in clinical practice \citep{cuenod23}; 
6/ improve the communication between clinicians by providing synthetic pictures.

In this segmentation framework, unfortunately, due to the diversity of tissues and their associated behaviors in response to the contrast agent injection, the number of regions/objects is unknown and clearly not predictable. Moreover, tissues could be filled, surrounded or interrupted by air or water but could also appear at several disconnected locations due to their geometry (e.g. colon). Thus, their number may vary from a few (a dozen) to many (several hundred). Therefore, a good segmentation should achieve an adaptive partitioning of the spatial domain into potentially disconnected functionally homogenous subsets/tissues. Here, adaptation is expected with respect to both the partition size and the subset shapes. 

\subsection{Overview of this paper}

To achieve the goals above, we propose a new method called {\it DCE-HiSET} where HiSET stands for {\it Hierarchical Segmentation using Equivalence Test}. 
Assuming constant noise level, DCE-HiSET employs a multiple equivalence test derived from a multi-resolution comparison in the time domain, known to be adaptive up to the unknown H\"older regularity of the unknown underlying signal using only $O(\log(n))$ comparisons. More specifically, given two TCs (or two averages of TCs), the equivalence test rejects when they are not sufficiently separated to be considered distinguishable. Thus, it allows to bind subsets that are functionally homogenous under this alternative.
Considering the $p$-value of the multiple equivalence test as a dissimilarity measure, DCE-HiSET is a bottom-up hierarchical clustering algorithm, consisting of two steps: one local and one global:  at each iteration the most functionally homogeneous pair of subsets are bound.\medskip\\
{\it In the local step} (growth of regions), starting from a partition made up of all voxels, only subsets that are spatially neighbors may be aggregated into larger clusters. 
\medskip\\
{\it In the global step} (merging regions), starting from the partition resulting from the local step, all clusters may be aggregated into larger clusters, regardless of whether they are neighbors or not.\medskip

These two steps ensure that {\it DCE-HiSET} will benefit from: 1/ the spatial regularity which can be expected inside tissues; 2/ a computational complexity which is lower than quadratic in the number of voxels; 3/ the opportunity to recover non-connected homogenous tissue.\medskip

DCE-HiSET depends on two intuitive parameters: $\alpha$, the significance level of the multiple equivalence test, controlling the probability of mistakenly binding two regions in the iterative process; $\delta$, the smallest expected homogeneity discrepancy between two (unobservable true) TCs. Empirically, the segmentation results show little sensitivity to $\alpha$; hence, it can be easily fixed as a meta-parameter. Parameter $\delta$ is a human interpretable parameter given it is meaningful on noiseless signals. It defines a kind of (multi-resolution) tube in which TCs are considered similar. Given $\alpha$ and $\delta$, both steps automatically stop through a proper control of the Type I error, providing an adaptive choice of the number of clusters. 
%

%


For an image sequence made up of functionally homogenous piecewise regions with signals on each piece that are different enough with respect to $n$, the number of images in the sequence, DCE-HiSET is theoretically proven to retrieve the true underlying partition with high probability.

For a real DCE sequence made of $n$ images, we propose a model of the observed intensities of the sequence, which depends only on one parameter $a$. DCE-HiSET is applied on the transformed sequence after a variance stabilization which ensure a constant noise level equal to 1. 
\medskip

\noindent {\bf Reading Guide:}  Hereafter, we review the literature related to image segmentation in the context of DCE image sequence. In Section \ref{sec:method}, we describe our statistical model and  objective (\S\ref{subsec:model}), the multiple equivalence test (\S\ref{subsec:test}) and the two-step clustering procedure with its theoretical properties (\S\ref{subsec:clustering}), along with the parameters involved (\S\ref{subsec:parameter}). In Section \ref{sec:exp}, we first introduce the material used for evaluation, including both synthetic and real DCE image sequences (\S\ref{sec:material}) and state-of-the-art competitors (\S\ref{sec:competitors}). Then, we present a comparison using the synthetic DCE image sequence (\S\ref{sec:evasyn}) and a study on real DCE image sequences, including parameter influence and model validation (\S\ref{sec:evareal}). Technical details and proofs are relegated to Appendices together with a table of our notations with their explanations.

\subsection{Prior approaches to DCE image sequence segmentation}\label{literature}

Region (object)-based segmentation has been investigated for detecting lesions \citep{DCEFCMKineticLesion2006,DCESupervoxel2016,DCESCKNNKmeans2014}, or for retrieving internal structure of organs using prior knowledge on the number of tissues in the organ of interest \citep{DCEGCGMM2009,DCEKmeansWavelet2012}. 
All these works use clustering-based image segmentation methods generally together with a supervised refinement step requiring a training dataset. 
Here, we focus only on an unsupervised approach that does not require a such training dataset. 
Moreover, we do not expect to have a prior knowledge of the number of tissues.  

TC representation in a DCE image sequence may be either parametric  \citep{DCESEAC2013, DCESCKNNKmeans2014,DCESupervoxel2016,DCESegSoftNull} or non-parametric  \citep{DCEKmeansWavelet2012}. The latter is known to provide a representation adaptive to the unknown TC regularity. This is not the case for parametric representations which are using a fixed number of descriptors.

Direct clustering of  the TCs had been proposed using $k$-means \citep{DCEMRIregist2011, DCEKmeansWavelet2012} or fuzzy c-means (FCM) \citep{DCEFCMKineticLesion2006}. It leads to segmenting the spatial domain by using the resulting TC labels. These approaches require an extra post-processing step (e.g. hole-filling) \citep{DCEFCMKineticLesion2006}. To incorporate spatial and time domain structures into one global procedure, two main types of segmentation have been proposed. The first consists of segmenting features obtained from binding spatial information (voxel coordinates) to the TCs \citep{MS2002}. The second considers the TCs to be distributed as a Gaussian mixture model (GMM) showing some regularity over the spatial domain ensured by a Markov Random Field (MRF) prior \citep{FCMHMRF2008,DCESegMRF2006}.

These previous approaches focused only on partial (or binary) segmentation. Despite being potentially adaptable to our purpose (complete segmentation of the spatial domain), to our knowledge, these methods have never been used for this objective.

Even for classic static images (grey, color, texture), few non-supervised methods have been proposed to solve the more complex problem of complete image segmentation. We review their main ideas and apply them when possible to DCE image sequence segmentation. These can be broadly classified into three categories: {\it model-based}, {\it graph-based} and {\it hybrid}. 

\paragraph[Model-based methods]{Model-based methods\eatpunct} describe the feature space as a mixture of models over the spatial domain. 
Minimizing the within-cluster distance, the segmentation is obtained by $k$-means \citep{DCEKmeansWavelet2012} or FCM \citep{DCEFCMKineticLesion2006}. 
Maximizing the posterior probability of a GMM with MRF prior, the segmentation is achieved by Expectation-Maximization (EM) \citep{EMMRF2003} or Bayesian sampling (MCMC) \citep{ISegMCMC2002}. $k$-means and FCM tend to find clusters with comparable shapes while the use of GMM allows clusters to have different shapes.
All these methods are strongly sensitive to their initialization step (seeds or initial partition) and suffer from the use of a pre-specified number of clusters. 
They tend to fail in presence of clusters with complex and unknown shapes, as previously mentioned in \cite{SelfTuningSpecClust2004}.

Mean shift (MS) \citep{MS2002} and quick shift (QS) \citep{QS2008} aim to find the modes of the distribution in the feature space, obtained by binding spatial coordinates and color information. Each individual voxel (or pixel) is assigned to a mode by minimizing a criterion computed on bound features. They do not need a predefined number of clusters; however they do need global kernel bandwidths for both time and spatial domain. Their choices are not automatic and require a strong expertise in the image to be segmented. Unfortunately, the use of global bandwidth cannot provide adaption in a general framework. 

\paragraph[Graph-based methods]{Graph-based methods\eatpunct} treat image segmentation as a graph-partitioning problem. 
The weighted graph is constructed from an image by considering voxels (or pixels) as nodes. The edges reflect the similarity between features, spatial distance or even both \citep{NC2000,SelfTuningSpecClust2004}. The partition is the result of a global minimization using the eigenvalues of a Laplacian matrix. Graph-based methods are able to handle feature space with more complex structures. 
However, they also require the knowledge of the number of clusters and some scale parameters to compute affinity \citep{NC2000}. 
\cite{SelfTuningSpecClust2004} addressed these two requirements through a self-tuning local scaling and a maximization of a cost function that depends on the number of clusters which is allowed to vary in a predefined set of values. 
Such approach has been used to segment DCE image sequences of prostate tumor \citep{DCESCKNNKmeans2014}, with the number of clusters ranging  between 1 and  5 only. 
In the context of tumor segmentation, \cite{DCESupervoxel2016} used a supervised step to discriminate between tissues/behaviors.

\paragraph[Hybrid methods]{Hybrid methods\eatpunct} consist of two steps -- one local and one global -- each derived from a model- or graph- based method described above \citep{ISegMCL2009,ISegMSNC2007}. These hybrid methods have been applied to DCE image sequence \citep{DCESupervoxel2016,DCEMSGC2014}. 
The local step over-segments the image into local homogenous clusters, called supervoxel (or superpixel). The global step merges these superpvoxels to obtain the final result.
Use of supervoxels is expected to reduce the sensitivity to noise and provide better segmentation, as mentioned in \cite{ISegMSNC2007}. 
Hybrid methods  share similar flaws with the methods they are based on in each step (see above).\medskip

\section{Method}\label{sec:method}

\subsection{Statistical observation model and objective}\label{subsec:model}
For DCE-MRI, at each voxel location $x$ on a finite grid $\mathcal{X}$ describing the image cross-section, the intensities $\Phi^x(t_j) \in \mathbf{R}_+$, $j=1,...,n$ are observed at the $n$ acquisition times, $t_1,t_2,...,t_n$. Up to the baseline, defined as the intensity before the arrival of the contrast agent, these intensities account for the amount of contrast agent particles within the voxel and within the acquisition delay. In queueing theory, it is usual to model the arrival increments as a Poisson distribution. Neglecting the baseline and assuming independent increments, namely independent arrivals of contrast agent particles, the amount of contrast agent particles $\Phi^x(t_j)$ within the voxel $x$ at time $t_j$ can also be assumed to be Poisson distributed. We denote by $\phi^x(t_j)$ its expectation such that $\Phi^x(t_j) \sim \mathcal P(\phi^x(t_j))$. Here, the Poisson distribution is used to model the signal enhancements due to the contrast agent injection. In this context, $\phi^x(t_j)$ may be considered large. This leads to the following Gaussian approximation of the distribution of $\Phi^x(t_j)$ together with its classical variance stabilization:
\begin{equation}
\Phi^x(t_j) \sim  \mathcal{N}(\phi^x(t_j),\phi^x(t_j)) \quad\text{ and }\quad 2\sqrt{\Phi^x(t_j)} - 2\sqrt{\phi^x(t_j)} \sim  \mathcal{N}(0,1).
\end{equation}
 
To relax our assumptions and provide more flexibility in the choice of the intensity distribution ({\it e.g.} heavier tails), we suppose that there exists $0<a<1$ such that 
\begin{equation}\label{model}
\Phi^x(t_j) \sim \mathcal{N}\left( \phi^x(t_j),\left[ \phi^x(t_j) \right]^{2-2a}\right),
\end{equation}
with a variance stabilized by the following:
\begin{equation}\label{varstab}
\frac{\left(\Phi^x(t_j)\right)^a}{a} - \frac{\left(\phi^x(t_j)\right)^a}{a} \sim  \mathcal{N}(0,1).
\end{equation}
In other words, denoting
\begin{equation}\label{a-model-1}
I^x(t_j):=\frac{\left(\Phi^x(t_j)\right)^a}{a} \quad\text{ and }\quad i^x(t_j):=\frac{\left(\phi^x(t_j)\right)^a}{a},
\end{equation}
we assume that the following model describes the transformed intensities
\begin{equation}\label{a-model-3}
 I^x(t_j) = i^x(t_j) + \eta^x_j, \qquad j = 1,\ldots,n \text{ and } x\in \mathcal X,
\end{equation}
where $\eta^x_j$ are standard Gaussian random variables independent with respect to time index $j$ (thanks to the production of one image at each time and to the significant delay between two consecutive image acquisitions). Spatial independence is much harder to justify due to the presence of spatial artifacts \citep{ArtifactMRI,CT}. Nevertheless, we make the following assumption in order to derive our mathematical construction.

\begin{ass}\label{ass3}
Random variables $\eta^x_j$ in \eqref{a-model-3} are standard Gaussian, independent with respect to both spatial location $x$ and time index $j$. 
\end{ass}

Furthermore, the {\it time curve} (TC) refers to the transformed version $i^x(.)$. Then, $I^x:=(I^x (t_1),\ldots,I^x(t_n))$ appears as a discretely observed noisy version of the unobservable true TC at time $t_1,\ldots,t_n$, that is of $i^x:=(i^x(t_1),\ldots,i^x(t_n))$.\medskip

Our statistical objective is to build a partition of $\mathcal{X}$ made of $\ell$ non-overlapping clusters (regions), $\mathcal{X} = C_1\cup C_2 \cup \ldots \cup C_{\ell}$ such that $x,y \in \mathcal{X}$ belong to the same cluster if and only if $i^x(.) = i^y(.)$. We propose to achieve this by answering the question ``are $i^x(.)$ and $i^y(.)$ equal or not?" from their discrete observations $I^x$ and $I^y$.\medskip

Integrating a more complex dependence in the spatial structure and extending our construction is certainly feasible but is far beyond the scope of this paper. Moreover thanks to the quality of our results and the study of the empirical residuals (see Section \ref{modelverif}), only small improvements may be expected from a more realistic modelization of the spatial dependency. Nonetheless, we can provide insights why these artifacts are having such small effects in the clustering. If they affect the full sequence on a small spatial domain, they will be integrated into the underlying unknown signals $i^x(.)$ for those $x$ in this domain and then be automatically compensated. Otherwise, when all images of the sequence are not affected in the same way\footnote{One can think of radial artifacts whose direction or center may vary from image to image, or of band artifacts that do not appear at the same coordinate}, the resulting spatial dependence varies from image to image and compensate by considering strategies based on the whole TC as in our construction.\medskip
 
The contrast agent is always injected a few seconds after acquisition starts, such that the baseline grey level may be estimated. Depending on their objective: tissue separation or analysis of the enhancements, radiologists or clinicians will focus on either the ``time intensity curve", which is made from the original intensities, or the ``time enhancement curve", obtained after the removal of the estimated baseline at each voxel. 

\subsection{Equivalence test and dissimilarity measure}\label{subsec:test}

Given a set of voxels $X\subset \mathcal X$, we consider the average discrete TCs
\begin{equation}\label{eq:barI}\bar I^X:= \frac{1}{|X|} \sum_{x\in X}I^x \quad\text{ and }\quad \bar i^X:= \frac{1}{|X|} \sum_{x\in X}i^x,\end{equation}
where $|X|$ is the cardinality of $X$.
Given another set $Y$ such that $X\cap Y=\emptyset$, we consider the scaling factor $\rho^2(X,Y):=|X|^{-1}+|Y|^{-1}$ and the normalized difference $D^{XY}:=(\bar I^X -\bar I^Y)\big /\rho(X,Y)$. Under Modeling assumption \ref{ass3}, $D^{XY}$ follows a multivariate Gaussian distribution $\mathcal N(d^{XY}, \textrm{Id}_n)$ with mean $d^{XY}:=(\bar{i}^X-\bar{i}^Y)\big / \rho(X,Y)$. 


The TCs of $X$ and  $Y$ will be considered as similar if their difference $d^{XY}$ is statistically not different from the zero vector. For this purpose, we consider an {\it equivalence test} of the hypotheses
\begin{equation}\label{multiequitest}
\mathcal{H}_0 : d^{XY} \neq \vec{0} \quad \textrm{versus} \quad \mathcal{H}_1 : d^{XY} = \vec{0},
\end{equation}
whose alternative (research) hypothesis is the null hypothesis of the conventional test of equality. The term {\it equivalence} highlights that equality is aimed to be shown\footnote{Equivalence testing derives from the need of pharmacists to prove that a drug-copy was making as well as the original drug up to a given tolerance. This is exactly our context here but for signals instead of drugs. In a classical testing construction with a given level of risks, even if all signals would be same a proportion close to the risk level will be declared different leading to many isolated voxels.}. 

As $d^{XY}$ belongs to $\mathbb R^n$, using one Gaussian equivalence test \citep{StatTesting} per time index would involve $n$ tests, leading to an important multiplicity problem\footnote{{a coordinate-by-coordinate equivalence test will compare each coordinate of the difference to 0, leading to $n$ comparisons each realized at a level $\alpha/n$ to ensure a global level $\alpha$ using a Bonferonni correction for simplicity. As a consequence the power of the test will be strongly reduced. Instead, the multi-resolution construction uses only $\log_2 n$ tests with a Bonferonni-corrected level $\alpha/\log_2 n$ ensuring a much smaller lost of power as $\log_2 128 \ll 128$. }}.  In order to control this, we modify the multiple test of \cite{BHL2003,BHL2005}, which use a dyadic decomposition of the time indexes into $\lfloor\log_2 n \rfloor$ time partitions. Our modifications, closely related to the Haar wavelet decomposition, ensure that the involved orthogonal projections are independent to control our clustering procedure.  
For $K=0,\ldots,K_0$ with $K_0:=\lfloor\log_2 n \rfloor-1$, the construction of the time partitions and the projection of $D^{XY}$ onto the $K$-th partition, denoted by $\mathrm{\Pi}_K D^{XY}$, are described in \ref{app1}. Starting from $D_{K_0+1}^{XY}=D^{XY}$, we consider the $K$-th {\it residual after projection}: $D_{K}^{XY}=\Pi_K D^{XY}-\Pi_{K-1} D^{XY}$, for $K=K_0+1,\ldots,0$ with $\Pi_{-1}D=0$. 
By construction, the $D_{K}^{XY}$ for $K=K_0,\ldots,0$ are orthogonal, which ensures their independence under Assumption \ref{ass3}, thanks to Cochran's theorem. Considering $K$-th {\it rescaled residual after projection} $\bar D_K = \Sigma_K^{-1/2} D_K$ where $\Sigma_K$ is the diagonal matrix defined by (\ref{eq:Sigma_K}), our test statistics are thereby $\|\bar D_K^{XY}\|^2_n$ instead of  $\|\Pi_K D^{XY}\|^2_n$ as in \citep{BHL2003,BHL2005}. Under Assumption \ref{ass3}, 
\[ \|\bar D_K^{XY}\|^2_n\sim\chi^2(2^{K-1},\| \bar d_K^{XY}\|^2_n) \quad\text{ for }\quad K=1,\ldots,K_0 \] 
and $\|\bar D_0^{XY}\|^2_n\sim\chi^2(1,\| \bar d_0^{XY}\|^2_n)$ where 
\begin{itemize}
\item $\|u\|_n$ denotes the Euclidian norm of vector $u$,
\item $\chi^2(\mu,\lambda)$ is the non-central chi-squared distribution with $\mu$ degrees of freedom and non-centrality parameter $\lambda$,
\item $\bar d_K^{XY}$ denotes the $K$-th rescaled residual after projection of $d^{XY}$.
\end{itemize}

We consider the  union-intersection {\it equivalence} test of the hypotheses
\begin{equation}\label{IUT}
\mathcal{H}_0 = \bigcup_{K=0}^{K_0} \mathcal{H}_0^K 
\quad \textrm{versus} \quad 
\mathcal{H}_1=\bigcap_{K=0}^{K_0} \mathcal{H}_1^K.
\end{equation}
In other words, $\mathcal{H}_1$ is accepted if and only if all $\mathcal{H}_1^K$ are. 

In this setting, ideally, one would like to chose $\mathcal{H}_0^K$ as $\| \bar d_K^{XY} \|^2_n \neq 0$ and $\mathcal{H}_1^K$ as  $\| \bar d_K^{XY} \|^2_n= 0$. Unfortunately, such hypotheses are not well separated statistically and one has to consider $\mathcal{H}_0^K$ of the form $\| \bar d_K^{XY} \|^2_n > n\delta_K^2$. The quantity $n\delta_K^2$ is called {\it equivalence margin}. It defines the discrepancy between two unobservable true TCs that we are ready to tolerate. 
 
%
As the following convergence holds
\[ \frac 1 n \| \bar d_K^{XY} \|^2_n \xrightarrow[n\rightarrow \infty]{} \frac 1 T \int_0^T \left(\bar d_K^{XY}(t)\right)^2 dt, \] 
it means that  $\mathcal H_1$ is rejected if the rescaled projected differences of the residuals shows some energy on any dyadic partition. In other words, two observations are considered to come from the same signal if one cannot detect a given level of energy in the rescaled residuals of their difference on any partition. 

Before we go any further, we would like to comment on the use of multiple test instead of one simple test such as one based on the Euclidian distance between discretely observed TCs. The multiple test has been shown \citep{BHL2003} to be optimal to detect departure from the zero signal without the knowledge of neither the departure nor its H\"older regularity. It is optimal in the sense that it achieves, up to a logarithmic factor, the same rate of detection as when the regularity of the departure is known, which is naturally an easier case. Moreover, any procedure which is based on only one partition may be shown to be sub-optimal. The simple test based on the Euclidian distance between discretely observed TCs is that of $\mathcal{H}_0^K$ against $\mathcal{H}_0^K$ with $K=0$, hence it is sub-optimal.

In order to define the $p$-value associated with our multiple equivalence test, we recall the following result.
\begin{prop}[\cite{IUtest}]
If $R_K$ is a rejection region at level $\alpha$, the union-intersection test with rejection
region $R=\bigcap_{K=0}^{K_0} R_K$ is  of level $\alpha$.
\end{prop}
%
Here, denoting $\Lambda_K$ a random variable following a shifted $\chi^2(2^{\bar K},n \delta^2_K)$ where $\bar K$ equals $K-1$ if $K>0$ and 1 otherwise, the $p$-value associated with the $K$-th test is
\begin{equation}\label{pvalue}
p_K(X,Y) := \mathbf{P} \left( \Lambda_K \leqslant \| \bar D_K^{XY} \|^2_n \right),
\end{equation}
and the following proposition holds.
\begin{cor} \label{prop1} The $p$-value of the union-intersection test defined by (\ref{IUT}) with rejection
region $R=\bigcap_{K=0}^{K_0} R_K$ is 
\begin{equation}\label{pIUT}
p(X,Y) := \max_K p_K(X,Y).
\end{equation}
\end{cor}
Despite our effort, we did not find this corollary in the literature, hence we provide its proof, which is reported in \ref{app:proof-prop1}. \hfill$\square$\medskip

The closer $p(X,Y)$ is from 0, the more similar the (average) TCs on $X$ and $Y$ are. Our hierarchical clustering is based on this observation and uses the $p(X,Y)$ as dissimilarity measure. \medskip

In the Gaussian setting, when the null is equality to zero on one given partition, it can be shown that the $p$-values are a monotonic transformation of an Euclidian distance. Hence, the rejection region can be defined by a distance. For the construction of \cite{BHL2003}, one can derive a multi-resolution distance as the maximum of monotonic transformations of the $p$-values. In our equivalence setting, where equality to zero is the alternative, we have not been able to derive a such distance and it is not clear if a monotonic transformation of the $p$-values can provide such a distance due to the presence of the equivalence margins which acts as thresholds.



\subsection{Clustering using equivalence test}\label{subsec:clustering}
A general setup of the clustering problem is to produce a partition of $\mathcal X$ into $\ell^*$ subsets $C_1\ldots C_{\ell^*}$, called clusters such that elements share similar properties in one cluster and have different properties from cluster to cluster. 
In the context of DCE image sequence, we expect voxels to have the same TC in one cluster and different TCs otherwise. 
Using the previously introduced equivalence test, here ``same" means that two voxels have the difference of their TCs under $\mathcal H_1$ so that the associated $p$-value is small. 

Our bottom-up hierarchical clustering algorithm is based on one main loop that aggregates two most similar clusters at each iteration. At iteration $\bar \ell$, assuming we have at hand a partition $\mathcal P^{\bar \ell}$ made of $\ell=|\mathcal X|-\bar\ell$ clusters $C_1^{\bar \ell},\ldots,C_\ell^{\bar \ell}$ together with 
indicators $\mathbbm 1_{s,s'}^{\bar \ell}$, $1\leqslant s,s' \leqslant \ell$, equal to 1 if $C_s^{\bar \ell}$ and $C_{s'}^{\bar \ell}$ are neighbors (defined hereafter) and 0 otherwise. 
Then the new partition is obtained by merging the two clusters indexed by $s_1$ and $s_2$ such that
\begin{equation}\label{minpair}
(s_1,s_2) = \arg\min_{\{(s,s')| 1\leqslant s,s' \leqslant \ell \text{ and } \mathbbm 1_{s,s'}^{\bar \ell}=1\}} p\left(C_s^{\bar \ell},C_{s'}^{\bar \ell}\right).
\end{equation}
We denote the {\it minimum dissimilarity} at iteration ${\bar\ell}$ by
\begin{equation}\label{mindiss}
p({\bar\ell}) := 
\min_{\{(s,s')| 1\leqslant s,s' \leqslant \ell \text{ and } \mathbbm 1_{s,s'}^{\bar \ell}=1\}} p\left(C_s^{\bar \ell},C_{s'}^{\bar \ell}\right).
\end{equation}
If the minimum may be achieved on more than one couple $(s,s')$, we chose the one with the smallest lexical order. 

The neighbors of a subset $C$, denoted $\mathcal V(C)$, is specified by the following: In the running partition, when two subsets are merged in one new cluster, the neighbors of the latter are those of both subsets. We now have to clarify the initialisation of each step.

The first step, called {\it local}, aims to take into account the spatial regularity existing in the images. It provides a preliminary partition made of connected clusters. It starts with the partition made of the $|\mathcal X|$ singletons: $\mathcal P^0:=\{\{x\}, x\in\mathcal X\}$. In this partition, two clusters are neighbors if and only if they are spatially connected on the (2D or 3D) grid $\mathcal X$. 

The second step, called {\it global}, aims to recover clusters with spatially disconnected sub-structures. It use as input the output partition of the first step. In this step, all clusters are neighbors of each other; hence all indicators are always set to one.

\subsubsection{Automatic selection of number of clusters}\label{asnc}

We already highlighted  that to achieve proper segmentation, the main concern is the protection of the functional homogeneity inside each cluster. Therefore, the choice of the number of clusters should correspond to the expected level of homogeneity, but not the other way around. In DCE image sequences,  when targeting a large area like the abdomen, the number of clusters can range from a dozen to a few thousands.

In order to automatically stop the iterations and to select a final partition, we introduce a {\it control function}, denoted by $c_\alpha(\ell)$, such that the iterations stop as soon as $p({\bar\ell})> c_\alpha(\ell)$. Starting from $\mathcal P^0$ with $p(0) \leqslant\ c_\alpha(|\mathcal X|)$, the local clustering recursively merges two clusters at each iteration until $p({\bar\ell})$ exceeds $c_\alpha(\ell)$, resulting in a partition $\mathcal P^{loc}$ made of $\ell^{loc}$ clusters. Then, the global clustering starts at iteration $|\mathcal X|-\ell^{loc}$ with a lower value for the minimum dissimilarity as the neighborhood structure is larger. Once again, the global clustering recursively merges two clusters at each iteration until $p({\bar\ell})$ exceeds $c_\alpha(\ell)$ again, resulting in a final partition $\mathcal 
P^*$ and producing a final number of clusters $\ell^*$. The pseudo coded algorithm is given in \ref{app:algorithm}.\medskip

The following definition specifies the notions of functional separation for two subsets and for a partition. \\[-12mm]
\begin{defi}\label{def1}\null\hfill\null
\begin{enumerate}
\item Two subsets $X$ and $Y$ of $\mathcal X$ are called ``{\it $\delta$-separated}" 
if their unobservable true TCs satisfy $\| \bar d_K^{XY} \|^2_n \geq n\delta^2$ for at least one value $K$, $0\leqslant K\leqslant K_0$.
\item $\mathcal X$ is a ``{\it $\delta$-partition of size $\ell$}'' if there exists a partition of $\mathcal X$ into $\delta$-separated subsets, $C_1,\ldots,C_\ell$. 
\end{enumerate}
\end{defi}

We have now the right setup to provide theoretical results for our procedure. The following two theorems control the probability of, first, stopping too late --- that is binding two $\delta$-separated subsets --- and, second, binding wrongly --- that is binding two non-separated subsets before two $\delta$-separated  ones. Thus, their corollary provides a control of the probability to recover the exact underlying partition.

\begin{theo}\label{theo1} {\bf - Stopping too late $=$ binding too much -}
Under Assumption \ref{ass3}, if $\mathcal X$ in a $\delta$-partition of size $\ell_0$, the probability that $\ell^*<\ell_0$ is lower than $\alpha>0$ as soon as
\begin{equation}\label{control}
c_\alpha(\ell) = \left(\frac{2\alpha}{\ell(\ell-1)} \right)^{1\over K_0+1}, \quad \text{ for every }1< \ell \leqslant |\mathcal X|.
\end{equation}
\end{theo}
The proof of Theorem \ref{theo1} is given in \ref{app:proof-theo1}.\hfill$\square$\medskip

\begin{theo}\label{theo2}  {\bf - Wrongly binding -}
Under Assumption \ref{ass3},  if $\mathcal X$ in a $\delta$-partition of size $\ell_0$, the probability that along the iterations two subsets of two $\delta$-separated clusters merge before any pair of non separated subsets is lower than $\beta_n=|\mathcal X|^3 (n/2)^{-\kappa}$ for $\kappa>1$ as soon as 
\[
n\delta_K^2\ge\max\left\{2^{1+\bar K/2};2(K+\kappa\log \frac n 2);1.57\sqrt{\kappa}\right\}
\]
which is ensured, when $n\geq 5$, by
\begin{equation}\label{the_max}
n\delta_K^2\geq \max\left\{\sqrt{n},2(1+\kappa\log 2) \log_2 \frac n 2\right\}.
\end{equation}

\end{theo}\medskip

A direct consequence of Theorems \ref{theo1} and \ref{theo2}  is the following result which controls the probability to achieve an exact recovery of the segmentation.
\begin{cor} {\bf - Exact partitioning -} Under Assumption \ref{ass3},  if $\mathcal X$ is a $\delta$-partition, it will be exactly recovered with probability larger than $1-\alpha-\beta_n$ with $\beta_n$ defined in Theorem \ref{theo2}.
\end{cor}
 
Let us point out that, for $n\leq 2000$ and $\kappa\geq 2$, the maximum in (\ref{the_max}) is always achieved at the second term. If $\delta_K$ is chosen to be 1, then for $n=100$ one can take $\kappa=11$. In this case for a DCE image sequence with $2D$-domain $\mathcal X$ of size $512\times512$ or equivalently with $3D$-domain $\mathcal X$ of size $25\times100\times100$ the probability $\beta_n$ of a wrong binding along all the iterations is less than $0.0037$. For a $2D$-domain of size $256\times256$, this probability drops to less than $5.8\times10^{-5}$. These values are realistic for DCE-MRI. For DCE-CT, usually $n$ is smaller around 30. Then one has to accept larger values of $\delta_K$ to keep having a large enough $\kappa$. Naturally, these values are provided by theoretical upper-bounds, in practice, $\beta_n$ may be much smaller. However, the latter theorem together with its corollary shows that for realistic DCE image sequence one can expect almost no error in the segmentation if tissue are sufficiently separated.

These two theorems do not provide an understanding of the benefit of using two steps in our algorithm. However, one can easily understand that the local step does not suffer from the combinatorial complexity of the global step, thanks to the neighborhood structure. Moreover, by  only aggregating neighbors, the local step also offers the opportunity to take into account a possible regularity existing over the domain of $\mathcal X$ between TCs.

The proof of Theorem \ref{theo2} is a direct consequence of the following lemma that controls the probability of making a mistake, at any step of the iteration, by binding two subsets $\delta$-separated before two subsets sharing the same unobservable true TC. 

\begin{lem}\label{lem2} {\bf - Binding $\delta$-separated before unseparated sets -}
The probability, that two $\delta$-separated clusters merge before two subsets of one cluster do, is less than $|\mathcal X|^2 (n/2)^{-\kappa}$ for $\kappa>1$ as soon as $n\delta_K^2$ satisfies one of the inequalities given in Theorem \ref{theo2}. \end{lem}
The proof of Lemma \ref{lem2} and Theorem \ref{theo2} are given in \ref{app:proof-theo2}. \hfill$\square$\medskip

The lines of proof of Lemma \ref{lem2} use that  $K_0$ ---the number of individual tests which define the multiple test--- is of order $\log_2 n$. Adapting this proof when the number of individual tests is 1 would lead to much larger $\delta_K$. Thus, the classical hierarchical clustering based on the Euclidean distance should lead to very poor segmentation, not being able to recover clusters when their TCs are not sufficiently separated. This is illustrated in \ref{app:chessboard} on a simulated image sequence.



\subsubsection{Adaptation to hierarchical clustering}
\begin{figure}[ht]
\begin{center}
\includegraphics[width=0.75\textwidth]{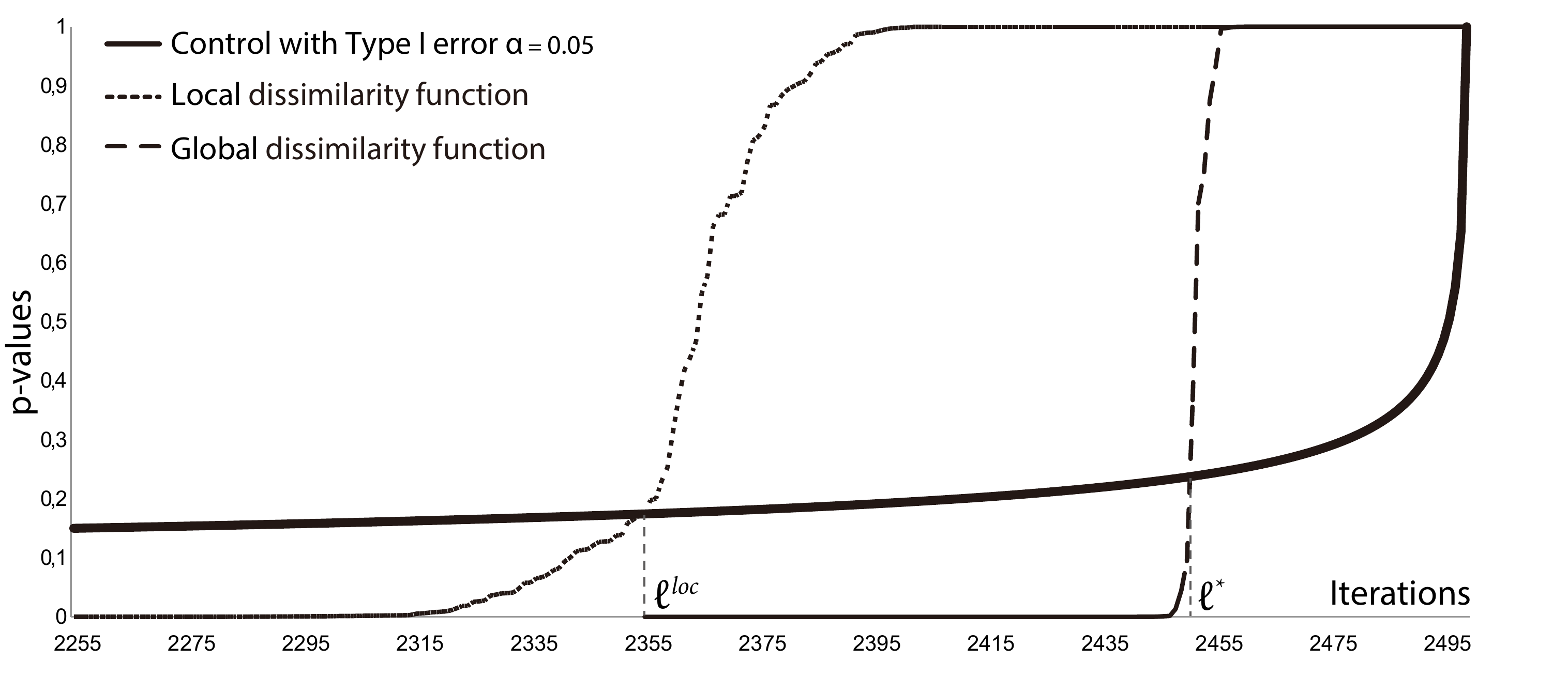}
\caption{\label{fig:dissfunc} {Selection of the number of clusters for both local and global clustering - Solid curve: the control function \eqref{control}. Dotted (resp. dashed) curve: the corrected minimum dissimilarity \eqref{cordisst} for the local (resp. global) clustering step. When the dotted curve reaches the solid one, at iteration $\ell^{loc}$, the local clustering stops. Using the resulting partition, the global clustering starts from $\ell^{loc}$ with the dashed curve. When the dashed curve reaches the solid one, at iteration $\ell^*$, the algorithm stops, providing the final partition. Minimum dissimilarity functions are shown even after they reach the control function to illustrate their typical behavior.}}
\end{center}
\end{figure}

In the conventional bottom-up hierarchical clustering \citep{HC2011}, the linkage function is defined to ensure that the minimum dissimilarity is non-decreasing along the iterations.
This is not guaranteed for HiSET so far. However, this could be fixed by considering the corrected dissimilarities $\bar p(.,.)$ defined as follows. Assuming $C_1$ and $C_2$ have been merged into $C$, then for $C'$ in $\mathcal V(C)$,
\begin{equation}\label{corrdiss}
\bar{p}( C,C') :=
\begin{cases}
\max\{ \bar{p}(C_1,C') ,p(C,C') \} &  \text{if } C' \in \mathcal{V}(C_1)\setminus \mathcal{V}(C_2);\\ 
\max\{ \bar{p}(C_2,C') ,p(C,C') \} &  \text{if } C' \in \mathcal{V}(C_2)\setminus \mathcal{V}(C_1);\\
\max\left\lbrace \min\{ \bar{p}(C_1,C')  ,\bar{p}(C_2,C') \} ,p(C,C') \right\rbrace & \qquad\text{otherwise.}
\end{cases}
\end{equation}
Thus, the minimum dissimilarity function (with correction) becomes
\begin{equation}\label{cordisst}
\bar{p}(\bar\ell) := 
\min_{\{(s,s')| 1\leqslant s,s' \leqslant \ell \text{ and } \mathbbm 1_{s,s'}^{\bar \ell}=1\}} \bar p\left(C_s^{\bar \ell},C_{s'}^{\bar \ell}\right).
\end{equation}
The selection of the number of clusters for both local and global clustering with a corrected minimum dissimilarity function is illustrated in Figure \ref{fig:dissfunc}. 

\subsection{Parameter interpretation} \label{subsec:parameter}
So far, we have introduced three parameters: 
1/ the ``model" factor $a$ in the variance stabilization transformation; 
2/ the equivalence margin $\delta$ defining the homogeneity discrepancy;
3/ the significance level $\alpha$ of the multiple equivalence test.
Parameter $a$ plays a role before the clustering starts, while $\delta$ and $\alpha$ are used during the clustering and only involved in the definition of, respectively, the minimum dissimilarity function \eqref{cordisst} and the control function \eqref{control}.
Their influence on the segmentation is studied in Section \ref{sec:evareal}.

\section{Experiments and results}\label{sec:exp}

\subsection{Material}\label{sec:material}
DCE-HiSET has been implemented using C++ code wrapped in R. Computation times for our examples range from a few seconds to a few dozens of seconds when $\delta$ is decreasing. DCE-HiSET has been evaluated, first on one synthetic DCE image sequence, and second on two real DCE-MR image sequences.

\begin{figure}[h!]
\centering
\includegraphics[width=0.3\textwidth]{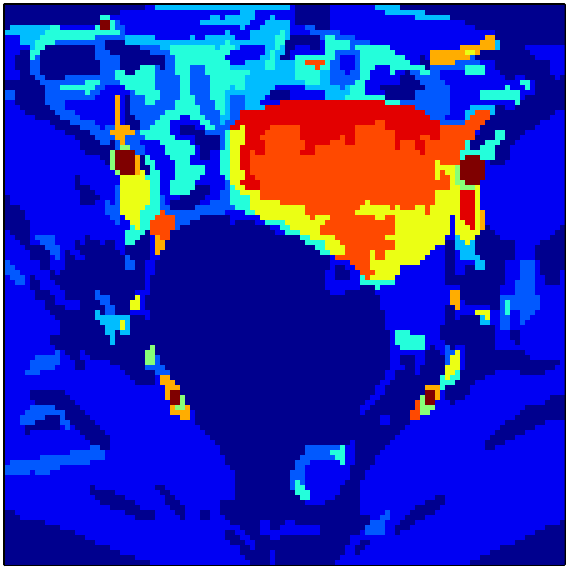} \qquad
\includegraphics[width=0.4\textwidth]{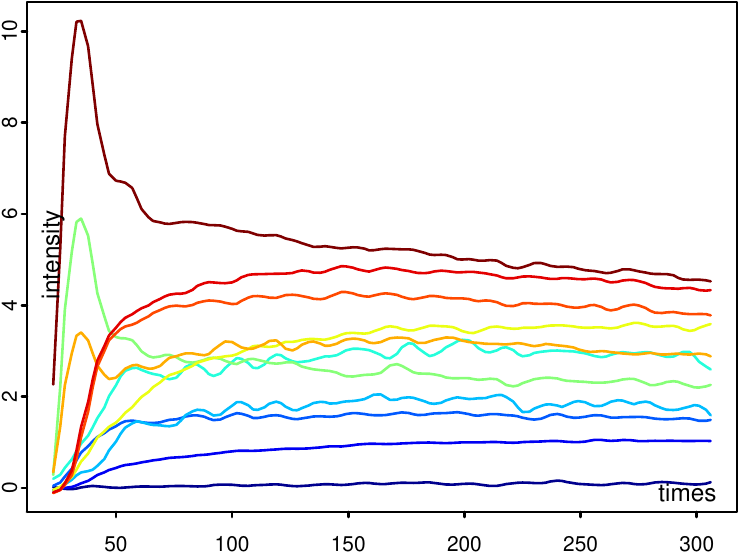}
\caption{Synthetic DCE image sequence: (left) The ground-truth segmentation of $\mathcal X$; (right) The true enhancement curves, $i^x(t)$, associated with the 11 clusters using corresponding colors.}
\label{fig:Sync}
\end{figure}

The synthetic sequence made by radiologists consists of 120 images on a grid $\mathcal X$ of 112x112 voxels. It contains eleven clusters of various sizes and shapes, representing real anatomical structures and their complexity, see Figure \ref{fig:Sync}. Independent standard Gaussian noises are added to the intensities at each time and at every voxel.

\begin{figure}[h!]
\centering
\hspace{2.5mm}\includegraphics[width=5.7cm]{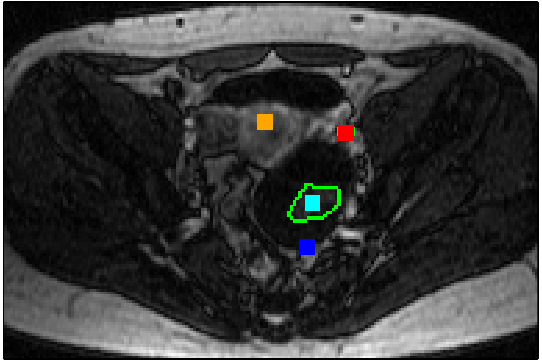} \hspace{2mm}
\includegraphics[width=5.7cm]{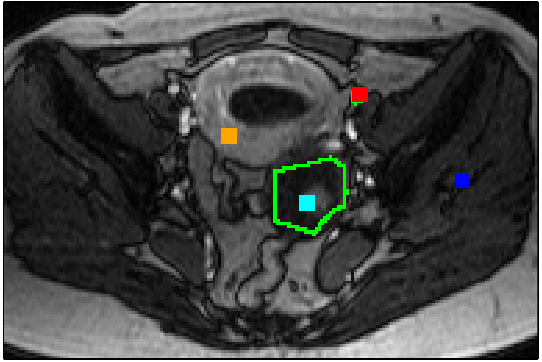}
 \\
\includegraphics[width=6cm]{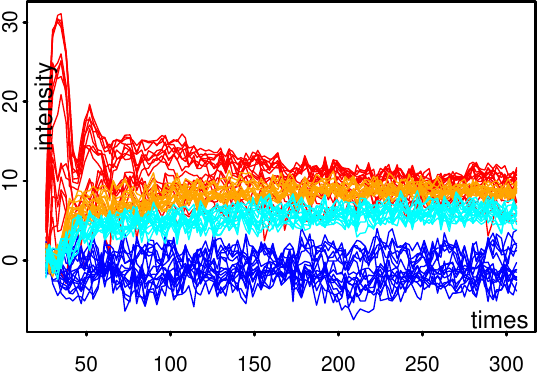}
\includegraphics[width=6cm]{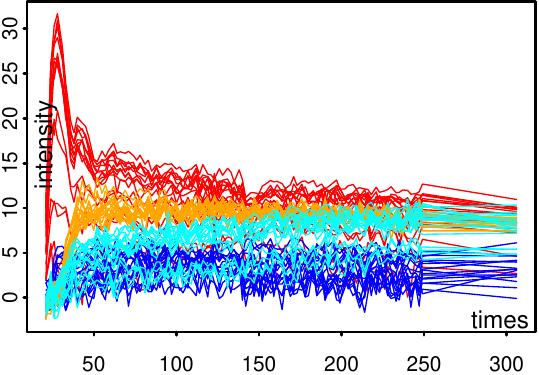}
\caption{DCE-MRI image sequences of two female pelvis with ovarian tumors - each column shows one sequence. Top: image obtained at time $t_{30}$ (\emph{after arterial phase}) with the tumor ROI (green) together with four 4x4 squared neighborhoods ({\emph red, cyan, orange and blue}), the red ones covering the iliac artery identified by the radiologist. Bottom (with corresponding colors): the sets of 16 time enhancement curves, $I^x(t_j)$, observed in the four squares after variance stabilization using $a=0.45$.}
\label{fig:ROIIS}
\end{figure}

The real DCE-MR image sequences of two female pelvis with ovarian tumors \citep{DCEMRIIS} consist of, respectively, 130 and 107 images on a grid $\mathcal X$ of 192x128 voxels, acquired over 305 seconds, see Figure \ref{fig:ROIIS}. 
With the delayed injection of the contrast agent bolus, their $n_0$ ($n_0=12$ and 10) first images show only the grey level baseline up to the noise level. In both sequences, a ROI around the tumor has been manually drawn by an experienced radiologist after acquisition. 



Each real DCE-MR image sequence is handled as follows: 
\vspace{-\topsep}
\begin{enumerate}
\setlength{\itemsep}{1pt} 
\addtolength{\itemindent}{-2mm}
\item The original intensity $\Phi^x(t_j)$, for each time $t_j$ and voxel $x$, is processed as in (\ref{a-model-1}) to obtain $I^x(t_j)$.
\item The baseline intensity, $b^x$, of each voxel $x$ is estimated by averaging the $n_0$ first intensities during the baseline phase, hence providing $\hat b^x$.
\item The baseline intensities are removed from the intensities to obtain the enhancements after baseline phase. For simplicity, the enhancements are also denoted $I^x(t_j)$:
\[I^x(t_j) \leftarrow \frac{I^x(t_j)-\hat b^x}{\sqrt{1+1/n_0}}, \quad \text{for }j=n_0+1,...,n.\]
These are Gaussians with a variance equal to 1 and are independent as soon as  Assumption \ref{ass3} holds for $I^x(t_j)$.
\item The image sequence is segmented from the $I^x(t_j)$ for $j=n_0+1,...,n$ and $x\in\mathcal X$.
\item After segmentation,  for each cluster $C$, the average intensity $\bar I^C$ is computed from \eqref{eq:barI} and normalized residuals are defined as
\[\hat \xi^x_j := \frac{I^x(t_j) - \bar I^C(t_j)}{\sqrt{1 - 1/|C|}}, \quad \text{for }j=n_0+1,...,n\]
for all $x$ in cluster $C$. These follow Gaussian distribution $\mathcal N(0,1)$ as soon as  Assumption \ref{ass3} holds for $I^x(t_j)$. \medskip
\end{enumerate}

The synthetic sequence directly provides  the $I^x(t_j)$ defined at the third step with $n_0=0$. 
Only the fourth and fifth steps are applied to this sequence. 

\subsection{Competitors}\label{sec:competitors}

To compare DCE-HiSET on the synthetic image, we considered the following 6 competitors from the 3 categories previously discussed in Section \ref{literature}:
\vspace{-0.8\topsep}
\begin{itemize}
\setlength{\itemsep}{-1pt} 
\addtolength{\itemindent}{-2mm}
\item model-based: $k$-means, HMRF-FCM \citep{FCMHMRF2008} and mean shift (MS) \citep{MS2002};
\item graph-based: normalized cut (NC) \citep{NC2000};
\item hybrid: MS followed by normalized cut (MS-NC) \citep{ISegMSNC2007}, SLIC \citep{SLIC2010,irving-2014} followed by normalized cut (SLIC-NC) \citep{DCESupervoxel2016}.
\end{itemize}
\vspace{-0.8\topsep}

Due to the noisy nature of image sequence, we employed NC only as the second step of the hybrid method such that the noise level had already been  reduced by the first step. 

\paragraph{Competitor details} 

All competitors, except for MS, require the number of clusters  as input parameter, and we implemented them according to the recommendations of their authors. 

$k$-means was used with Euclidean distance between TCs. It is well known that $k$-means is highly sensitive to initialization. Hence, for each value of $k$, we ran $k$-means with 250 different random initializations and picked  the one with the best value of the objective function.

For other competitors, a PCA decomposition of the features has been used first with 3 eigenvectors then 6. Three was enough to explain 95\% of the variance; 6 corresponds to the number of tests used in our multiple equivalence test for the sake of fairness. We observe no significant difference between these two choices and report results only with six eigenvectors.

HMRF-FCM requires an input partition and is highly sensitive to its choice. It uses the $k$-means result as input and then optimizes a regularized objective function of a FCM-type rather than a EM-type \citep{EMMRF2003}, which has been proven to be effective for vector-valued image \citep{FCMHMRF2008}.  

MS-NC, NC uses the output of MS as input; therefore it can only reduce the number of clusters from the MS step. 

By design SLIC provides an over-segmentation which is used as the initialization of NC in SLIC-NC.\\\medskip

When reviewing the literature, we realized that not only were some methods not available as a package but that some details on the parameter selection were also missing. To compensate for the latter, we used an Oracle approach: for each method and each parameter, the range of the latter had been manually chosen to cover under- and over-fitting. Inside these ranges, the parameters were chosen to provide the best value of the objective function. This approach led to the following parameter ranges.
\begin{itemize}
\setlength{\itemsep}{-1pt} 
\addtolength{\itemindent}{-2mm}
\item when required, the number of clusters was allowed to vary from 7 to 15.
\item the multiplicative constant $\varepsilon$ for the entropy penalty controlling the fuzziness in HMRF-FCM was allowed to vary from 0.2 to 4 by increments of 0.1.
\item MS and MS-NC use a temporal and a spatial bandwidth, denoted respectively $bw_t$ and $bw_s$. We have let $bw_t$ vary from 0.03 to 0.1 by increments of 0.01 and $r_{ts}:=bw_t/bw_s$ from 3 to 15 by increments of 0.5.
\item  SLIC uses a size and a compactness of supervoxels, denoted respectively $sv_s$ and $sv_c$. We have let $sv_s$ vary from 3 to 10 by step of 0.5 and $sv_c$ from 0 to 0.05 by increments of 0.005.
\end{itemize}

\paragraph{Competitor implementations}

All competitors and DCE-HiSET were run on the same laptop (macbook pro with i5 core 2.5Ghz with 16Gb of RAM) to provides comparable running times.

We used the $k$-means implementation provided in $R$\footnote{\url{https://cran.r-project.org/}} by the function {\tt kmeans}. Depending on the number of clusters, the computation time ranged from 10 to 90 minutes for the 250 runs.

We were not able to find available code for HMRF-FCM. Therefore, we used a $C$-implementation wrapped in $R$ with computational time varying from 1.5 to 10 minutes given the initial partition, depending on both the convergence rate (the number of iterations needed to reach convergence) and the number of clusters.

We used the implementation of MS\footnote{\url{http://a-asvadi.ir/}} in Matlab\footnote{\url{http://www.mathworks.com}} with computation time varying from 0.3 seconds to 33 seconds, depending on the values of bandwidths. 
The $C$-implementation of NC\footnote{\url{http://www.timotheecour.com/}} wrapped in Matlab resulted in computation time always less than 1 second. The same held for SLIC\footnote{\url{http://www.vlfeat.org/}} using its $C$-implementation wrapped in Matlab with computation time depending only on the supervoxel size. For MS-NC, the computation time depends on the gap between the  number of clusters resulting from MS and the one expected by NC.

\subsection{Evaluation on the synthetic sequence}\label{sec:evasyn}

\paragraph{Evaluation criterion}
For the synthetic sequence, the accuracy of segmentation results has been measured by the Fowlkes-Mallows Index (FM) \citep{FMIndex}, 
\[\text{FM} = \frac{N_{11}}{\sqrt{(N_{11} + N_{10})(N_{11} + N_{01})}},\]
where the number of voxel pairs is denoted $N_{11}$ when both voxels are classified into the same cluster in both partitions; $N_{10}$ when they are in the same cluster of the first partition but two different clusters in the second partition; $N_{01}$ when they are in two different clusters in the first partition but the same cluster in the second partition.
This index lies between 0 and 1: the closer it is to 1, the more similar  the segmentations. It accounts for the proportion of pairs of voxels which are in the same cluster in both partitions with respect to the geometric average of being at least in a same cluster for one partition. That is the proportion of well classified pairs by both methods with respect to the average number of pairs which could be well classified by at least one method. Here the use of the geometrical mean instead of the classical arithmetical mean provides more weight to the largest value. As already noted, keeping the cluster homogeneous with respect to the time intensity curve is the top concern, especially as the number of clusters cannot be considered as known. In this context, Fowlkes-Mallows Index provides a good trade-off for measuring how partitions show the same type of results or not. It is indeed a better end-point than the number of clusters given that 1/ in our context, it is unknown; 2/ two partitions can have the same number of clusters and show strong variations in the construction of their clusters.

To take into account the effect of cluster size on the accuracy measure, we also consider a weighted version, wFM. In this case, a pair of voxels $(x_1,x_2)$ is counted for $w_1\cdot w_2$ instead of 1, where, for $i=1,2$, $w_i=\frac{|\mathcal X|}{|C|}$ assuming $x_i$ belongs to cluster $C$ in the first partition. Errors in this weighted version are balanced with respect to the cluster size.


In order to provide a representation of the difference between two partitions $(C_1,\ldots,C_{\ell})$ and  $(D_1,\ldots,D_{\ell'})$, we also compute the {\it error-map} as the indicator function of the set
\[ \bigcup_{i=1}^{\ell} \left(C_i \setminus D_{j_i}\right) \quad \cup \quad \bigcup_{j=1}^{\ell'} \left(D_j \setminus C_{i_j}\right) \]
where $i_j = \arg\max_i |D_j \cap C_i|$ and $j_i = \arg\max_j |C_i \cap D_j|$. \medskip

While the Fowlkes-Mallows index accounts for pairs of similar voxels which are or not in the same cluster of the second partition, the error map counts for voxels of each cluster $A$ (in the first partition) which are not in the {\it best} cluster $B$ (in the second partition), where the best cluster $B$ is defined as to cover the most of $A$ and vice-versa. Hence, together they offer two different points-of-view on how close two partitions are. We considered more indices; however they all showed similar behavior to one of the two described above. Therefore, for sake of simplicity, we choose the ones with intuitive interpretation.

\begin{figure}[h!]
\centering
\includegraphics[width=0.4\textwidth]{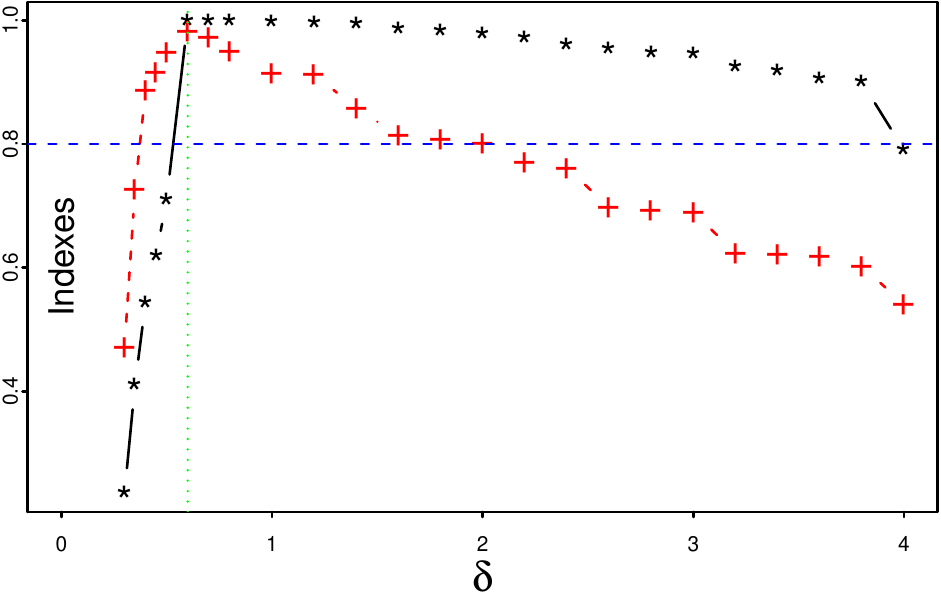} \hspace{5mm}
\includegraphics[width=0.4\textwidth]{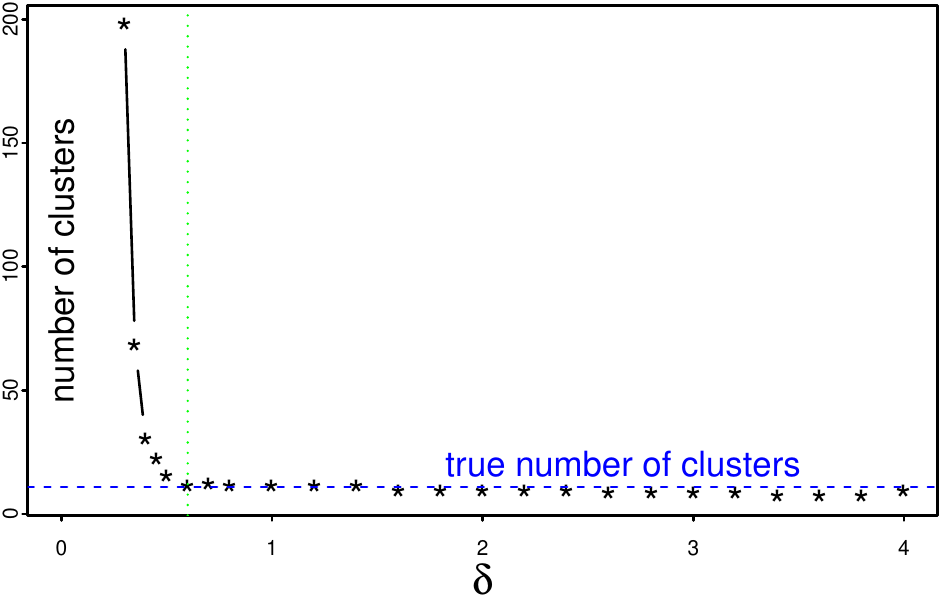}
\caption{Synthetic image sequence segmentation using DCE-HiSET when $\delta$ varies: 
(Left) Fowlkes-Mallows Index (black stars) and its weighted version (red crosses) - (Right) number of clusters. Result with best indexes is achieved at $\delta=0.6$ (green dashed line).}
\label{fig:SyncRes}
\end{figure}

\paragraph{Behavior of DCE-HiSET}
We used DCE-HiSET with a fixed value of $\alpha=0.001$ and various values of $\delta$ to segment the synthetic image sequence. The effect of the latter on the segmentation accuracy and on the number of clusters is illustrated in Figure~\ref{fig:SyncRes}.  We observed that the FM and wFM indexes achieve their maximum for the same value, denoted $\delta^*$. Moreover, they remain stable and higher than 80\% in a large range of $\delta$ from 0.5 to 2. The choice of a fix value for $\alpha$ is justified further in Section \ref{sec:param-influence} as this parameter has only a small influence on the final result with respect to $\delta$.  Indeed, it only controls  the probability that the final partition is too small and not the size of the partition itself. 

\begin{figure}[h!]
\centering
\includegraphics[width=0.3\textwidth]{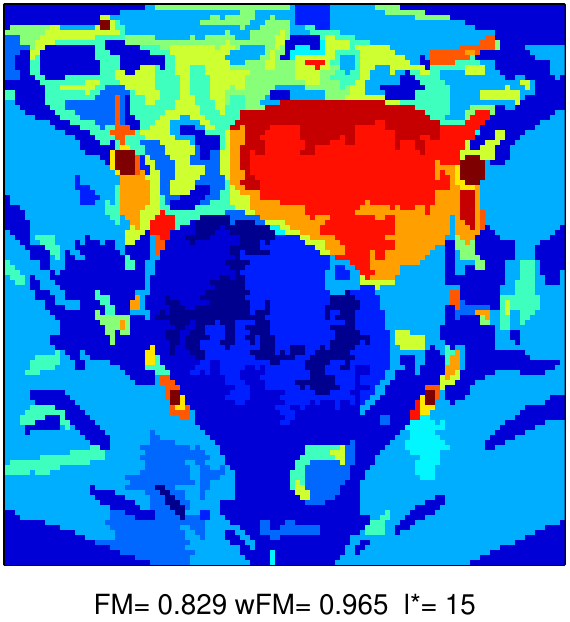} 
\includegraphics[width=0.3\textwidth]{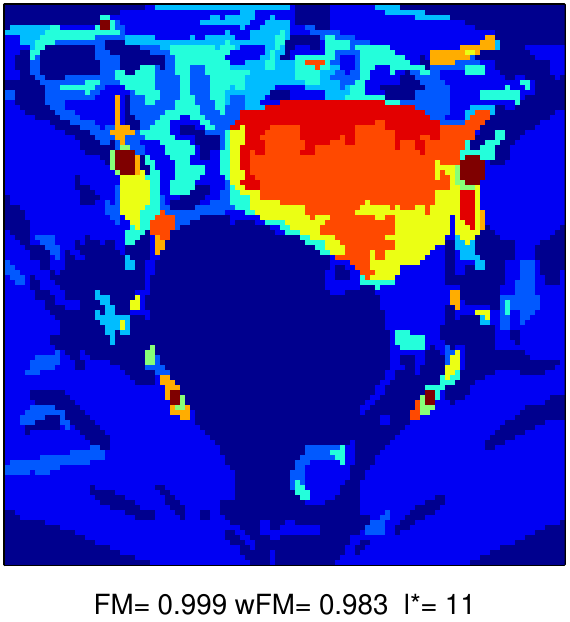} 
\includegraphics[width=0.3\textwidth]{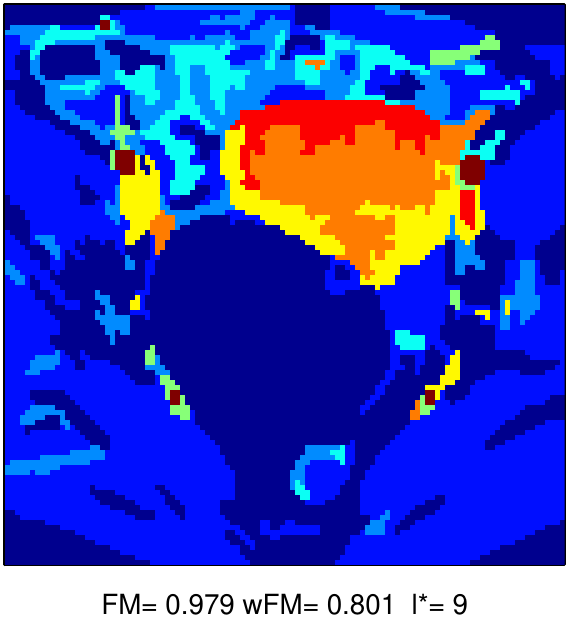}
\caption{Segmentation results of DCE-HiSET of synthetic image sequence with $\delta$ equals to 0.5 (left), $\delta^*=0.6$ (middle) and 2.0 (right) when $\alpha=0.001$.}
\label{fig:SyncRes-delta}
\end{figure}

Figure~\ref{fig:SyncRes-delta} shows the segmentation results with 3 values of $\delta$, including $\delta^*$. Clearly, small values of $\delta$ result in over-segmentation ($\ell^*$ too large): large homogeneous regions are split into sub-regions with irregular borders; however, these sub-regions remain large. In contrast, large values of $\delta$ enforce under-segmentation: true clusters start to merge; however the border geometry is not changed, ensuring a good recognition of most of the structures. Nevertheless, as one can expect when thinking of bias-variance trade-off, it is clearly more risky to use too small values than too large ones. 

For a fully automatic procedure with automatic selection of $\delta$ following the slope heuristic \citep{SlopeHeuristics2012,ProbTheory2007}, the fast decrease of the number of clusters when $\delta$ increases may provide an automatic choice of $\delta$: one can detect the value $\delta_0$ of $\delta$ where the relative slope, defined as $(\ell^*(\delta^-)-\ell^*(\delta^+))/\ell^*(\delta^-)$, is less than a fixed value and then define the optimal $\delta$ as $\delta^*=2\delta_0$.  \medskip

\begin{figure}[h!]
\centering
\includegraphics[width=0.45\textwidth]{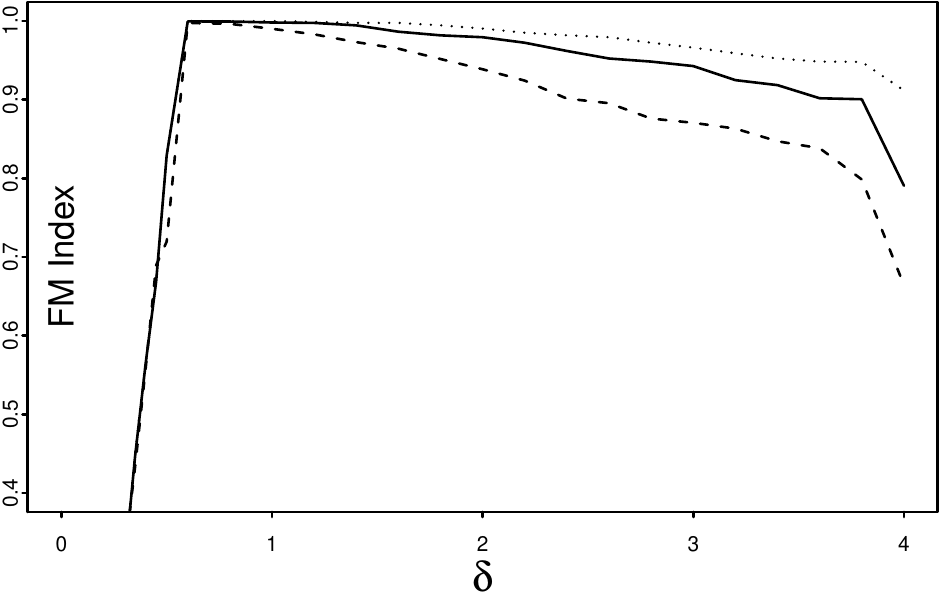} \hspace{5mm}
\includegraphics[width=0.45\textwidth]{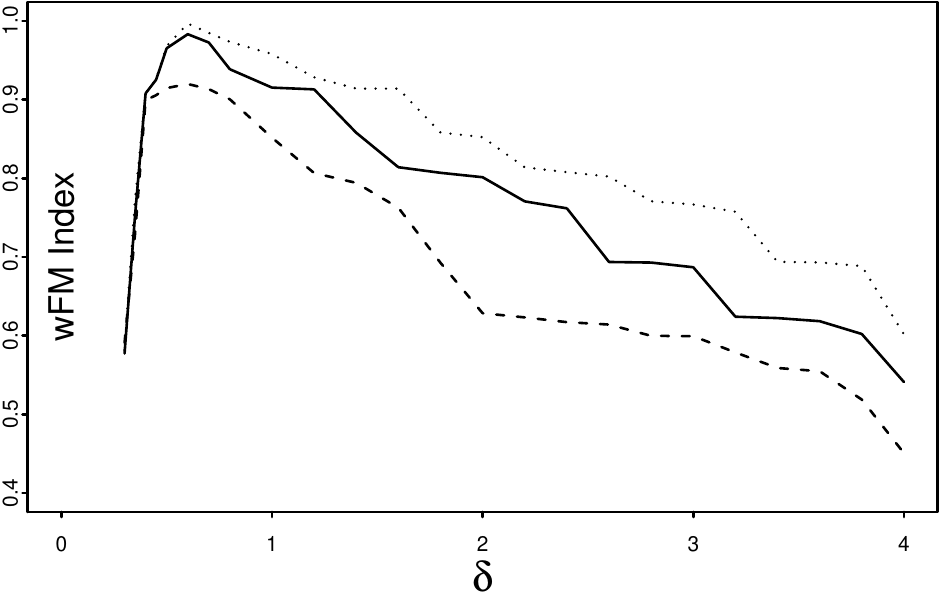}
\caption{FM index (left) and weighted FM index (right) when true enhancement  are multiplied by 4/3 (dotted), 1 (solid) and 2/3 (dashed).}
\label{fig:FM versus delta}
\end{figure}

\begin{figure}[h!]
\centering
\includegraphics[width=0.3\textwidth]{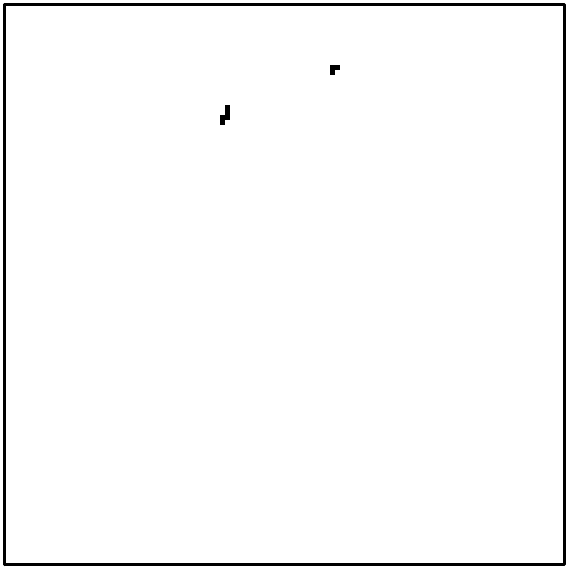}
\includegraphics[width=0.3\textwidth]{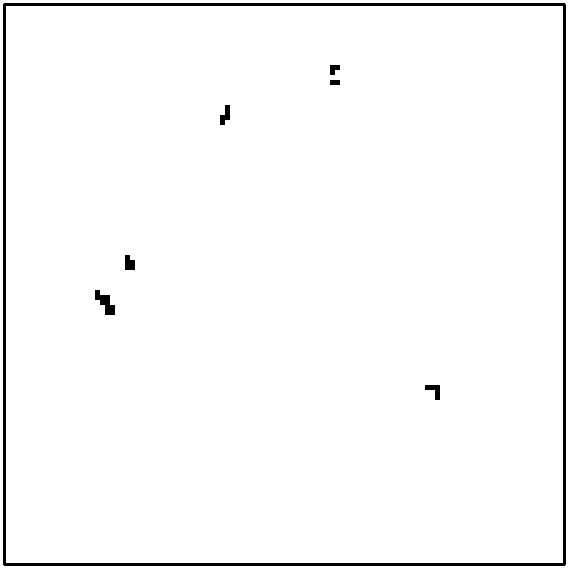}
\includegraphics[width=0.3\textwidth]{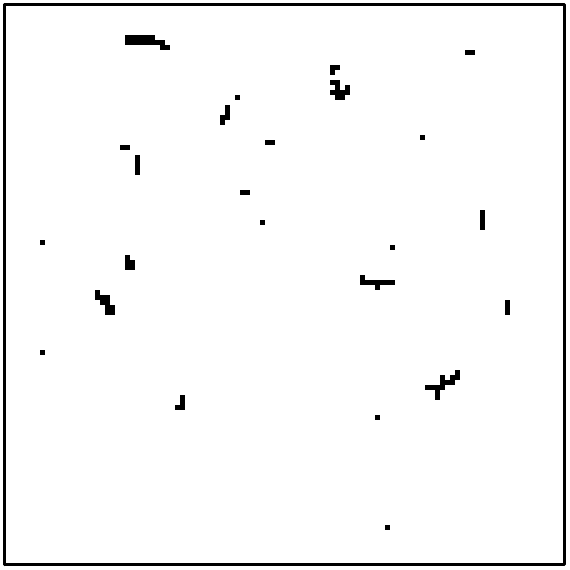}
\caption{Error maps for wFM-optimal value of $\delta$ when true enhancement  are multiplied by 4/3 (left), 1 (middle) and 2/3 (right).}
\label{fig:error map snr}
\end{figure}

In order to study the effect of TC separations on the segmentation results and the stability of DCE-HiSET with respect to the parameter $\delta$, we have multiplied the true enhancement curves by 2/3 and 4/3, while the noise level was kept fixed with a standard deviation equal to 1. In this setting, both the separation distance and the SNR are affected by the multiplicative factor and the segmentation is all the harder the smaller the multiplicative factor becomes. Figure~\ref{fig:FM versus delta} shows that both FM and wFM indexes achieve optimal value for the same value $\delta^*$, indicating that the key factor is indeed the ratio between the separation distance and the SNR. Clearly,  the harder the segmentation, the smaller is the range around the optimal value when relaxing the index maximization. The error maps corresponding to $\delta^*$ are given in Figure~\ref{fig:error map snr}.

\paragraph{Comparison with competitors}
Information and Oracle performances of all competitors together with DCE-HiSET are summarized in Table \ref{tab:comp}. 
\begin{table}[h!]
\begin{center}
{\footnotesize
\begin{tabular}{| l | c | c | c | c |}
\hline 
Method		& 	Parameter	& 	Range by Step 		& 	Highest FM 		&	Highest wFM 				\\ \hline
$k$-means	& 	$k$	  	&	7--15 by 1		&	0.993 ($k=8$)   	&      0.827 ($k=10$)				\\ \hline
\multirow{2}{*}{HMRF-FCM}	
			&	$k$		&	7--15 by 1	& \multirow{2}{*}{0.993 $\left(\begin{array}{c}k=8\\\varepsilon=0.5\end{array}\right)$}	   															& \multirow{2}{*}{0.827 $\left(\begin{array}{c}k=10\\\varepsilon=1\end{array}\right)$}			\\ \cline{2-3}
			& $\varepsilon$	& 	0.2--4 by 0.5		& 					&		     					\\ \hline
\multirow{2}{*}{MS}	
			&	$bw_{t}$	&	0.03--0.1 by 0.01	&	0.994 			& 	0.919 		     			\\ \cline{2-3}
			& 	$r_{ts}$	& 	3--14 by 0.5		& \multicolumn{2}{|c|}{($bw_{t}=0.06$, $r_{ts}=11$)}			\\ \hline
\multirow{3}{*}{MS-NC}	
			& 	$k$		& 	7--15 by 1		& 	0.993 			& 	0.852		     			\\ \cline{2-3}
			&	$bw_{t}$	&	0.03--0.1 by 0.01		& \multicolumn{2}{|c|}{($k=10$, $bw_{t}=0.06$, $r_{ts}=11$)}	\\ \cline{2-3}
			& 	$r_{ts}$	& 	3--14 by 0.5		& 					&	   						\\ \hline
\multirow{3}{*}{SLIC-NC}	
			& 	$k$		& 	7--15 by 1		& 	0.983			&	0.821     					\\ \cline{2-3}
			&	$sv_{s}$	&	3--10 by 0.5		& \multicolumn{2}{|c|}{($k=12$, $sv_{s}=3$, $sv_{c}=0$)}		\\ \cline{2-3}
			& 	$sv_{c}$	& 	0--0.05 by 0.005		& 					&	  				   		\\ \hline
DCE-HiSET  	&     $\delta$	&	0.2--4 by 0.1		& {\bf \color{red} 0.999} ($\delta=0.6$)   		
										& {\bf \color{red} 0.983} ($\delta=0.6$)	     				\\ 
\hline
\end{tabular}
}
\caption{Fowlkes-Mallows index with its weighted version: for each competitor, the input parameters are provided with their associated used range. The Oracle, given in parentheses, achieves the highest value of the index over the parameter ranges. }
\label{tab:comp}
\end{center}
\end{table}

\begin{figure}[h!]
\centering
\includegraphics[width=0.3\textwidth]{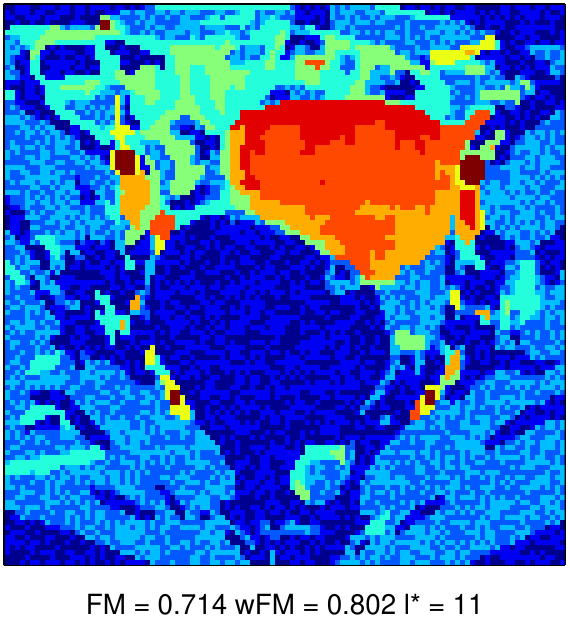} 
\includegraphics[width=0.3\textwidth]{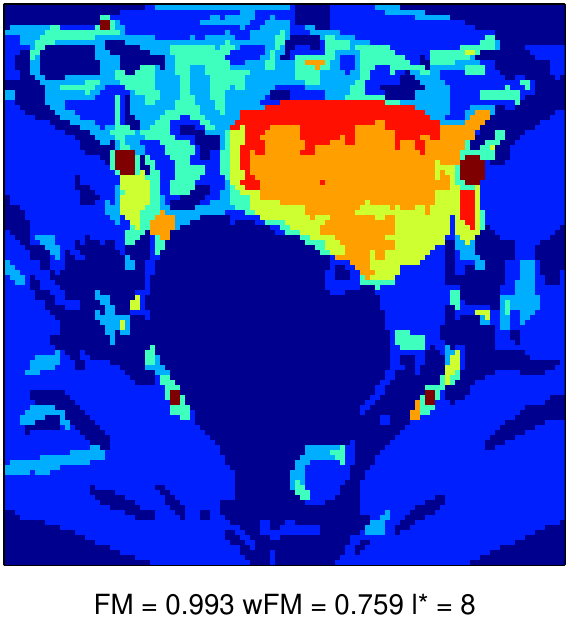} 
\includegraphics[width=0.3\textwidth]{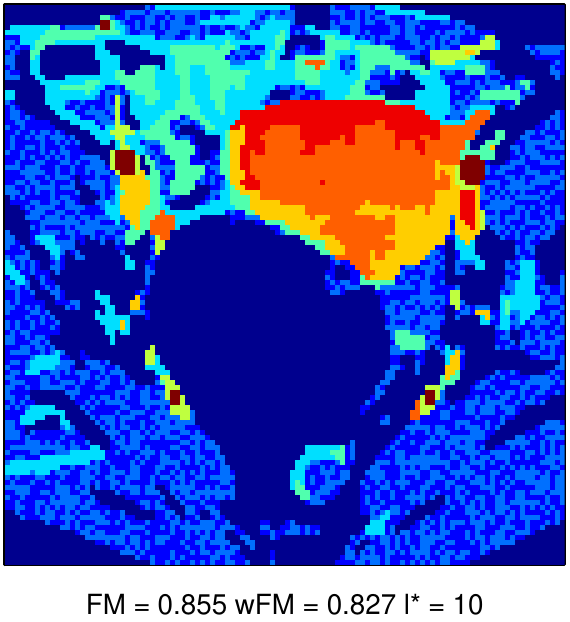} 
\includegraphics[width=0.3\textwidth]{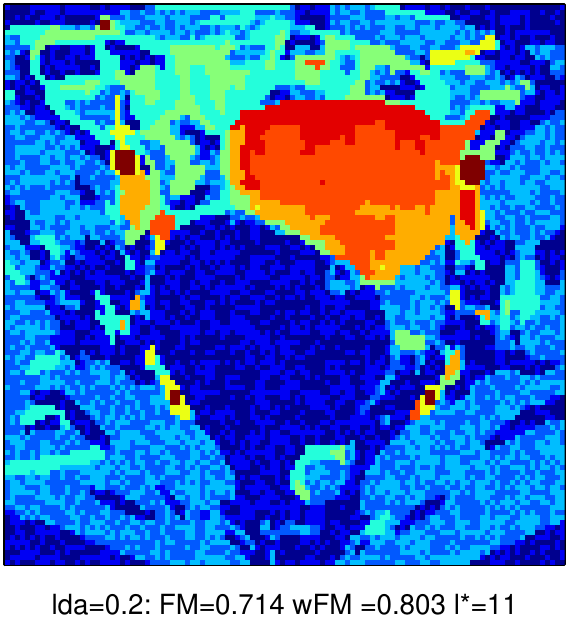} 
\includegraphics[width=0.3\textwidth]{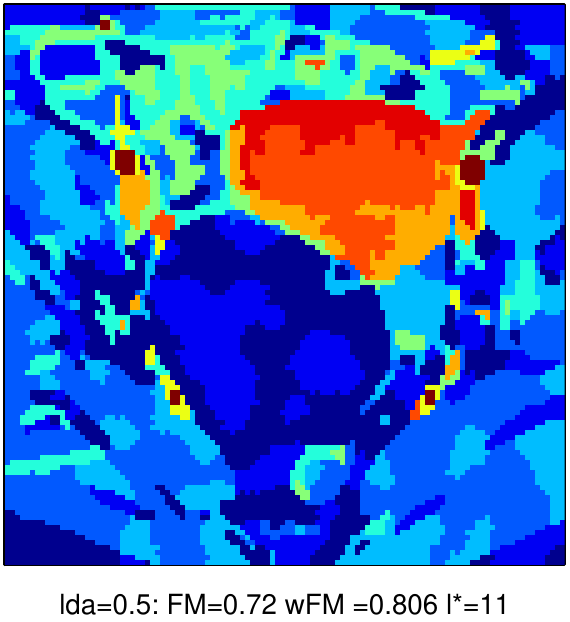} 
\includegraphics[width=0.3\textwidth]{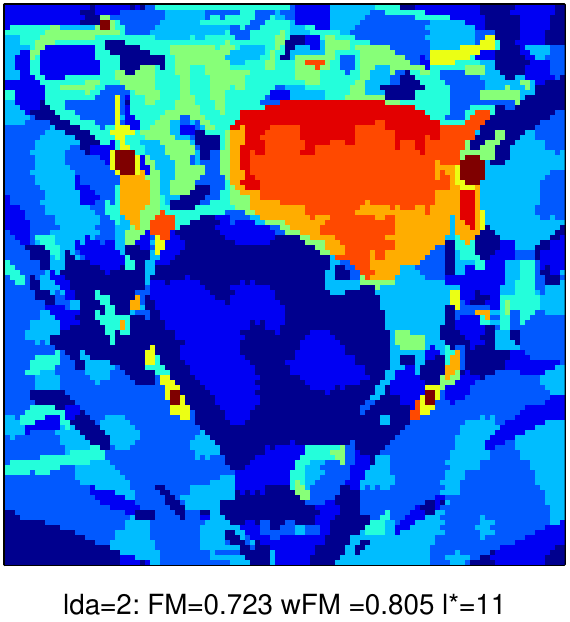} 
\caption{Oracle results -
Top: Best result of $k$-means among 250 runs with 
true number of clusters  and with: $k=11$ (left);  highest FM index (middle); highest wFM index (right).
Bottom: Results of HMRF-FCM initialized by the result of $k$-means with $k=11$ and with: $\varepsilon=0.2$ (left); $\varepsilon=0.5$ (middle); $\varepsilon=2$ (right).}
\label{fig:kmeans&HMRFFCM}
\end{figure}

For $k$-means, Figure~\ref{fig:kmeans&HMRFFCM} shows the result when $k=\ell^*=11$ that is the true number of clusters, together with the best FM and wFM values achieved when $k$ varies. The weaknesses of $k$-means are clearly visible: on one hand, having two or more initial centers in one true cluster (see $\ell^*=10$ or 11 in Fig. \ref{fig:kmeans&HMRFFCM}) results in pulverized clusters (known as the {\it pepper and salt} effect); on the other hand, when no center reaches small clusters, the latter are artificially merged into other clusters, leading to a high FM but low wFM (see $\ell^*=8$ in Fig. \ref{fig:kmeans&HMRFFCM}).

\begin{figure}[t!]
\centering
\includegraphics[width=0.3\textwidth]{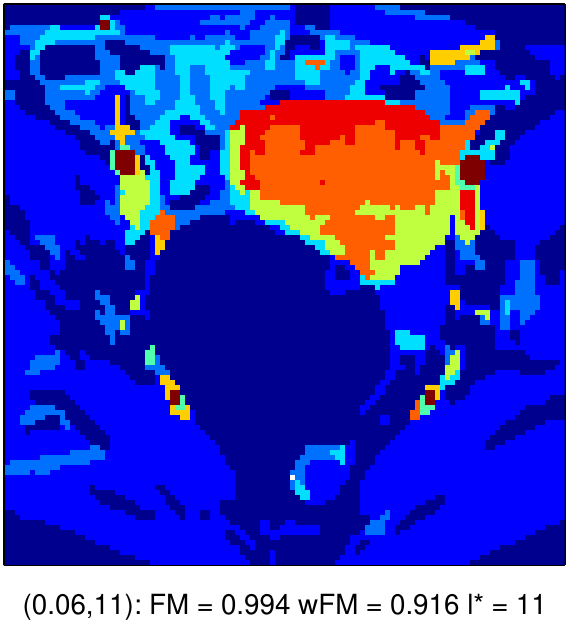}
\includegraphics[width=0.3\textwidth]{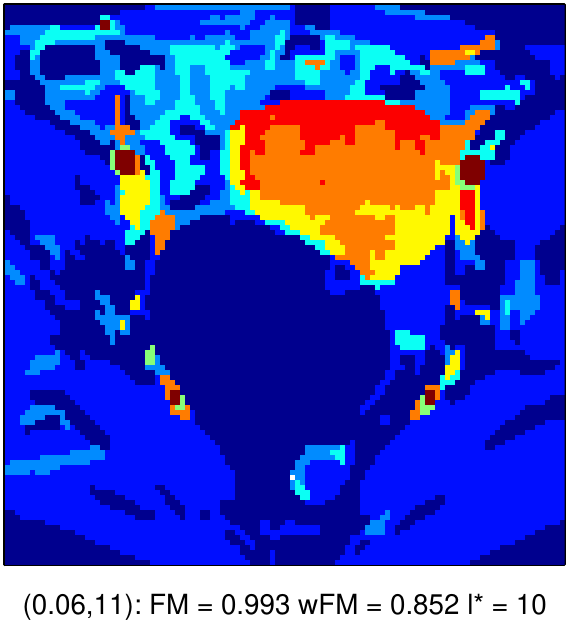}
\includegraphics[width=0.3\textwidth]{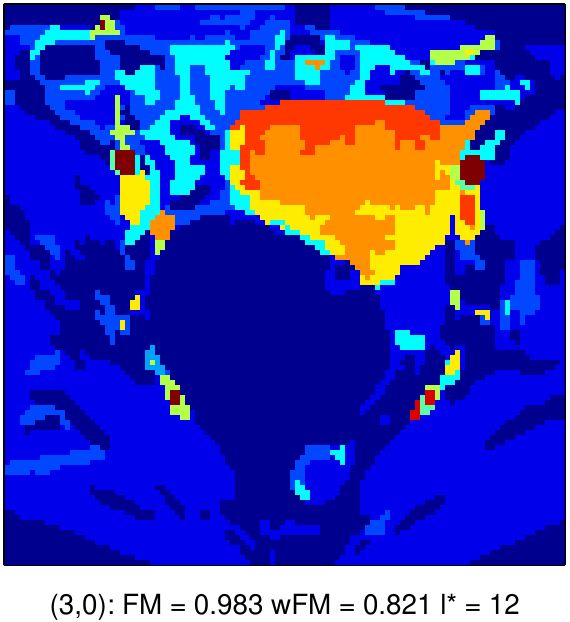}
\caption{Oracle results -  MS (left); MS-NC (middle); SLIC-NC (right).}
\label{fig:MSSLICNC}
\end{figure}

If the initial partition given by $k$-means has no pulverized cluster, HMRF-FCM barely improves the segmentation of $k$-means. Otherwise, it regularizes the pulverized clusters by grouping their voxels into a ``panther texture" when $\varepsilon$ becomes large enough (see Fig. \ref{fig:kmeans&HMRFFCM}).
However (not presented here), if one cluster has already been  split in the initial partition provided by $k$-means, HMRF-FCM is unable to fix it. Thus, the increase of FM and wFM induced by HMRF-FCM is very small for a given $k$. 
Moreover, the best FM and wFM indexes, when $k$ and $\varepsilon$ vary, are unchanged with respect to $k$-means.
For each of MS, MS-NC and SLIC-NC (see Fig.~\ref{fig:MSSLICNC}), as for DCE-HiSET, the best results for both indexes are achieved at same parameter sets, showing a good and expected property of stability with respect to the size of the clusters. 
MS recovered the right number of clusters. 
High FM (0.994) and relatively low wFM (0.916) indicate that mistakes have been made in small clusters. 
MS-NC recovers only 10 clusters with worse indexes than for MS alone. Indeed, two clusters resulting from the best result of MS have merged during the NC step. SLIC-NC has neither pulverized issues, as with $k$-means, nor mistakes in small clusters, as with MS, thanks to the first step with SLIC. 
However, due to the supervoxel size limitation in SLIC, voxels in small clusters may be spread in the surrounding large clusters and/or narrow parts of clusters tend to be thicker.

Despite their weakness in term of indexes, MS, MS-NC and SLIC-NC demonstrate the ability to recover regions highly consistent to the ground-truth segmentation as DCE-HiSET does.

\begin{figure}[h!]
\centering
\includegraphics[width=0.24\textwidth]{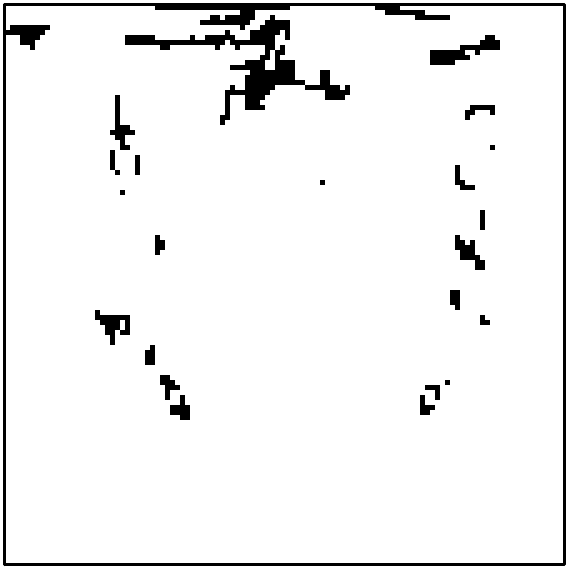}
\includegraphics[width=0.24\textwidth]{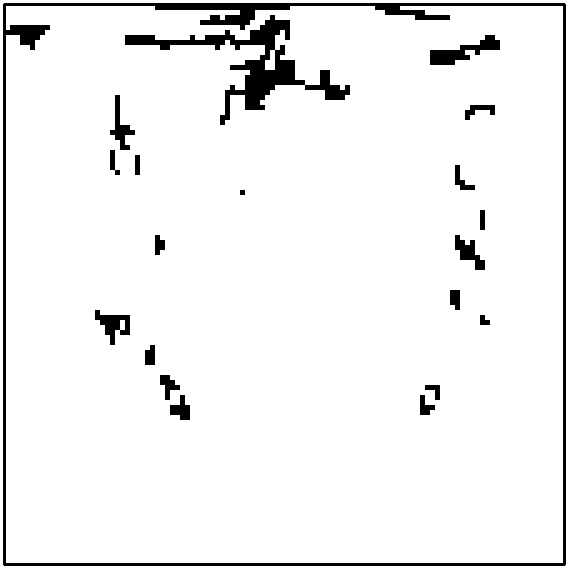}
\includegraphics[width=0.24\textwidth]{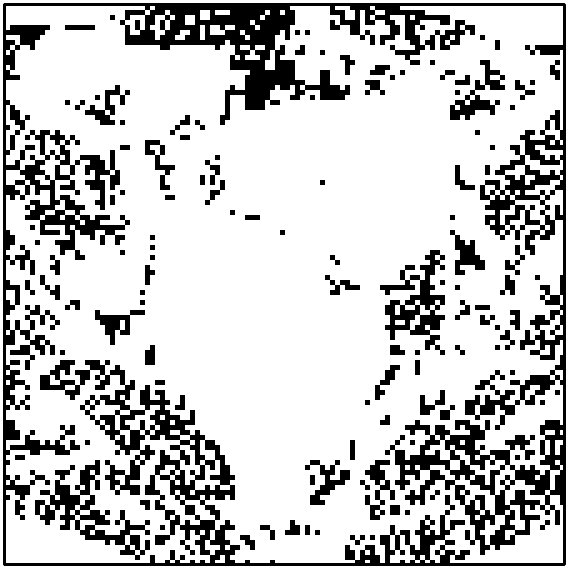} 
\includegraphics[width=0.24\textwidth]{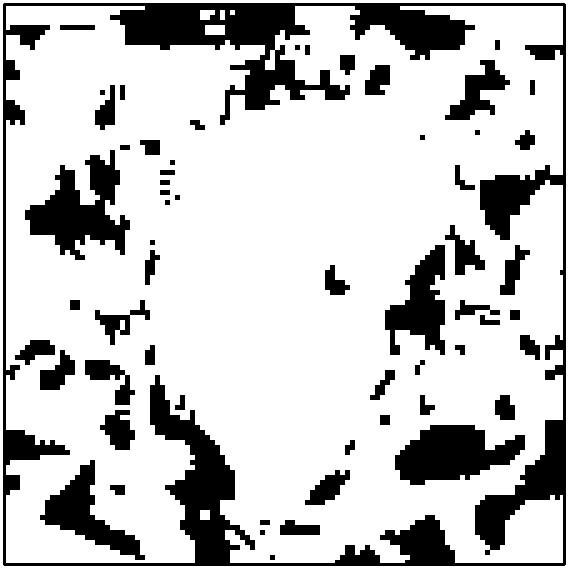} \\
\includegraphics[width=0.24\textwidth]{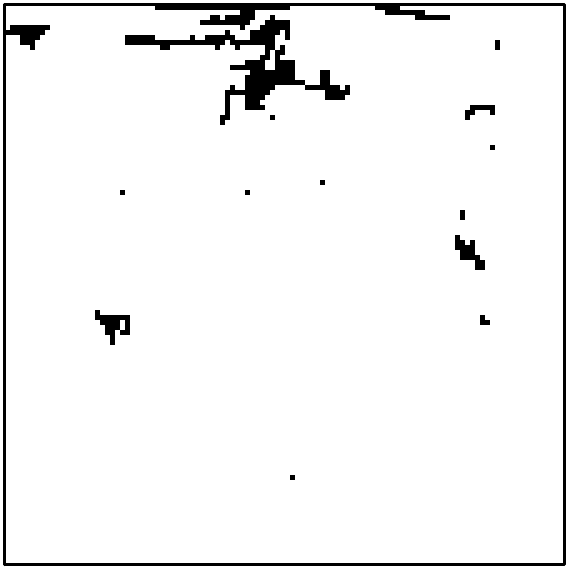}
\includegraphics[width=0.24\textwidth]{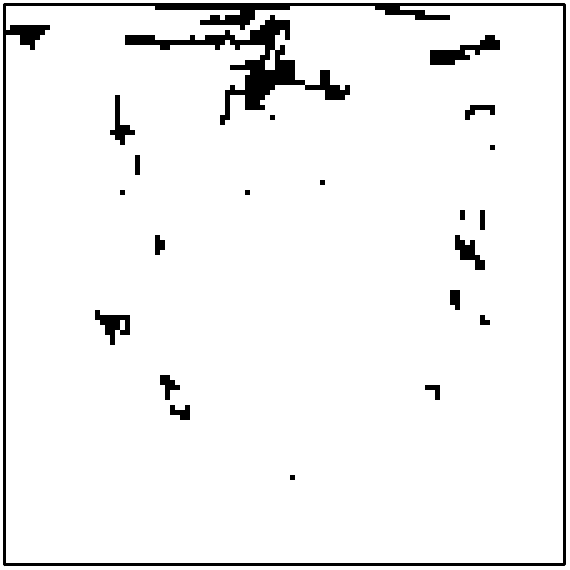}
\includegraphics[width=0.24\textwidth]{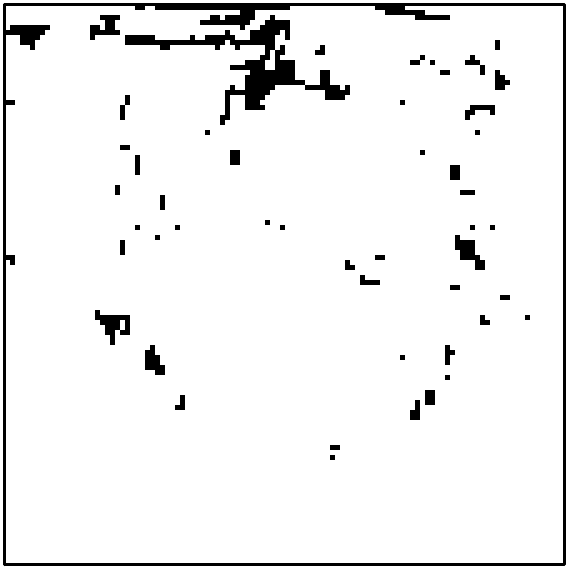}
\includegraphics[width=0.24\textwidth]{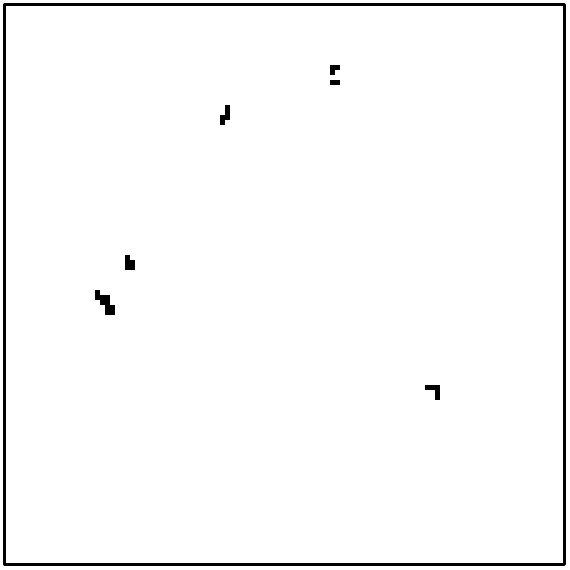}
\caption{Error maps - From left to right: (top) $k$-means and HMRF-FCM with highest FM; $k$-means and HMRF-FCM with highest wFM; (bottom) best results of MS, MS-NC, SLIC-NC (here the best results are achieved on same parameter sets for both indexes) and the (adaptive) result of DCE-HiSET. }
\label{fig:discrep}
\end{figure}



Figure \ref{fig:discrep} shows the error maps of the Oracle of each competitor together with the error map of DCE-HiSET to compare how segmentation errors (misclassified voxels) are distributed.

From this study of the synthetic sequence, we see clearly that DCE-HiSET outperforms all competitors with respect to both FM and wFM indexes while using only a single parameter. DCE-HiSET  shows moreover a very stable behavior. The error maps show clearly that all competitors have trouble with small and/or complex clusters while DCE-HiSET makes fewer errors in these areas, thanks to the multi-resolution comparison. Moreover, when the SNR decreases, the segmentation errors of DCE-HiSET expand regularly all over the image.

\subsection{Experiment on real DCE-MRI sequences}\label{sec:evareal}

\begin{figure}[h!]
\centering
\includegraphics[height=3.5cm]{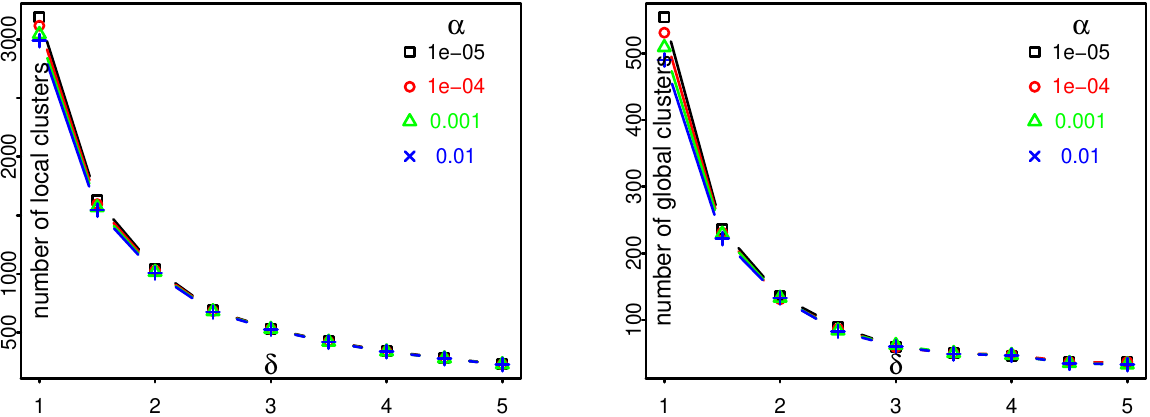}
\caption{Number of clusters when $\alpha$ and $\delta$ vary for (\emph{left}) local and (\emph{right}) global steps.}
\label{fig:Para}
\end{figure}

\subsubsection{Parameter influence}\label{sec:param-influence} Figure~\ref{fig:Para} shows the influence of parameters $\alpha$ and $\delta$ on the segmentation size. Clearly $\alpha$ has a negligible effect on the partition size compared to that of $\delta$, in both local and global steps. Hence, we fixed $\alpha=0.001$ for the rest of this study. Of note, we also observed that the cluster shape is not much influenced by $\alpha$ either; $\alpha$ only controls the probability of stopping too late and not how clusters are built along the iterations.

\subsubsection{Model verification and selection of parameter $a$}\label{modelverif}

\begin{figure}[b!]
\centering 
\includegraphics[width=\textwidth]{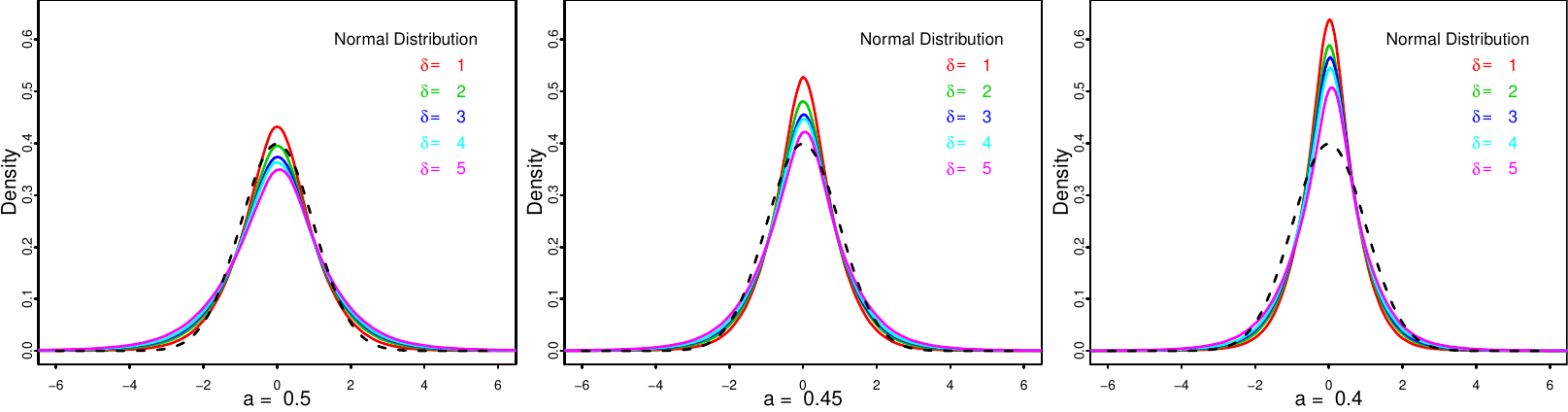}
\caption{Normalized residuals - estimated densities for $\delta\in\{1,2,3,4,5\}$ and normal density as reference (black and dashed) when using a variance stabilization with $a$=0.5 (left); 0.45 (middle); 0.4 (right).}
\label{DenresMalin}
\end{figure}

In order to justify our model given by equations (\ref{model}-\ref{a-model-3}), we studied the distribution of the normalized residuals $\hat \xi^x_j$ when $\delta$ and $a$ vary (see Fig.~\ref{DenresMalin}). We observed that the distribution tails of $\hat \xi^x_j$ are heavier than those of a Gaussian distribution when $a$ is large (Poisson's case: $a=0.5$) and are lighter when $a$ is too small ($a=0.4$); this remains true for all values of $\delta$. For the intermediate value $a=0.45$, the distribution shows a behavior close to Gaussian (or at least sub-Gaussian {\it i.e.} having only Gaussian tails). In this case, $\delta$ becomes the proper tuning parameter to obtain a residual distribution close to (sub-) Gaussian. Let us point out that, thanks to the use of tests, only tails are of interest. Extension of our construction to the sub-Gaussian case would require many more theoretical developments beyond the scope of this paper.  Similarly, adaptation with respect to $a$ or improving the model to get a better variance stabilization would be an objective of future investigation. Nevertheless, the quality of our empirical experiments shows that the benefit of this will be probably small. \medskip

We continue this study with a fixed value $a=0.45$.

\subsubsection{Segmentation results}

\begin{figure}[b!]
\centering
\rotatebox{90}{\scriptsize \qquad\qquad Patient 1}
\includegraphics[height=3.5cm]{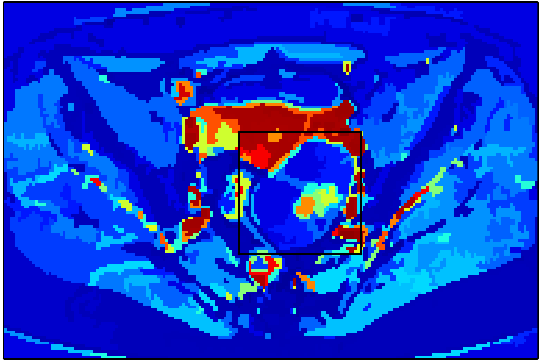}\hspace{3mm}
\rotatebox{90}{\scriptsize \qquad\qquad Patient 2}
\includegraphics[height=3.5cm]{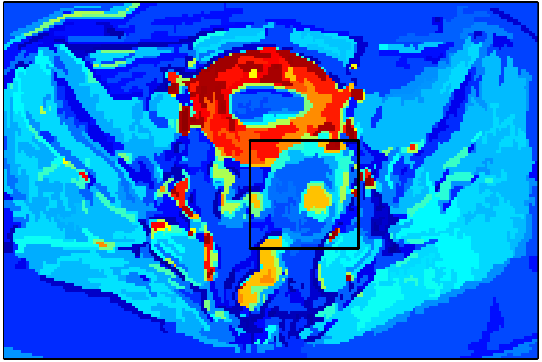}
\caption{Segmentation results of the two real DCE-MR image sequences using $\delta=3$.}
\label{fig:SegRes}
\end{figure}

\begin{figure}[h!]
\centering
\rotatebox{90}{\scriptsize \qquad\qquad Patient 1}
\includegraphics[height=3.3cm]{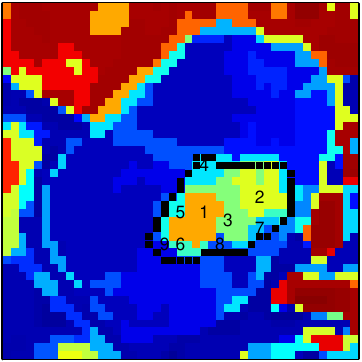}\quad
\includegraphics[height=3.3cm]{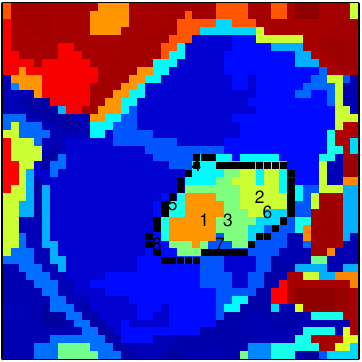}\quad
\includegraphics[height=3.3cm]{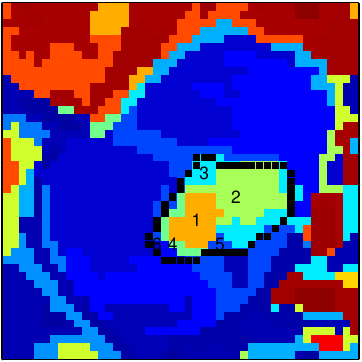}\\\medskip
\rotatebox{90}{\scriptsize \qquad\qquad Patient 2}
\includegraphics[height=3.3cm]{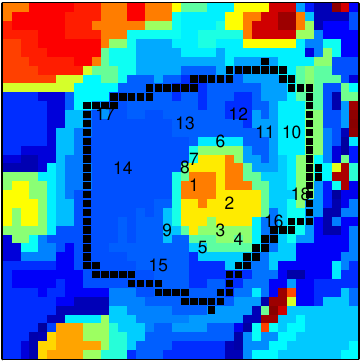}\quad
\includegraphics[height=3.3cm]{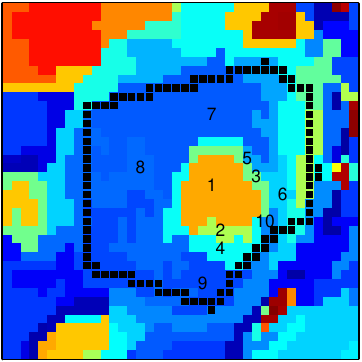}\quad
\includegraphics[height=3.3cm]{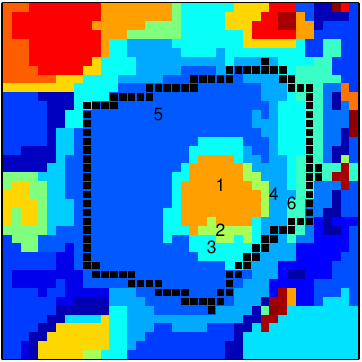}
\caption{Squared zoom in the DCE-MR image sequences with $\delta$ equals 2 (left); 3 (middle); 4 (right). 
The manually segmented ROI appears in black and clusters inside are numbered. The first sequence  (top) shows from 9 to 6 clusters. The second  (bottom) from 18 to 6.  
}
\label{fig:SegROI}
\end{figure}

\begin{figure}[h!]
\centering
\rotatebox{90}{\scriptsize \qquad\quad Patient 1}
\includegraphics[height=2.5cm]{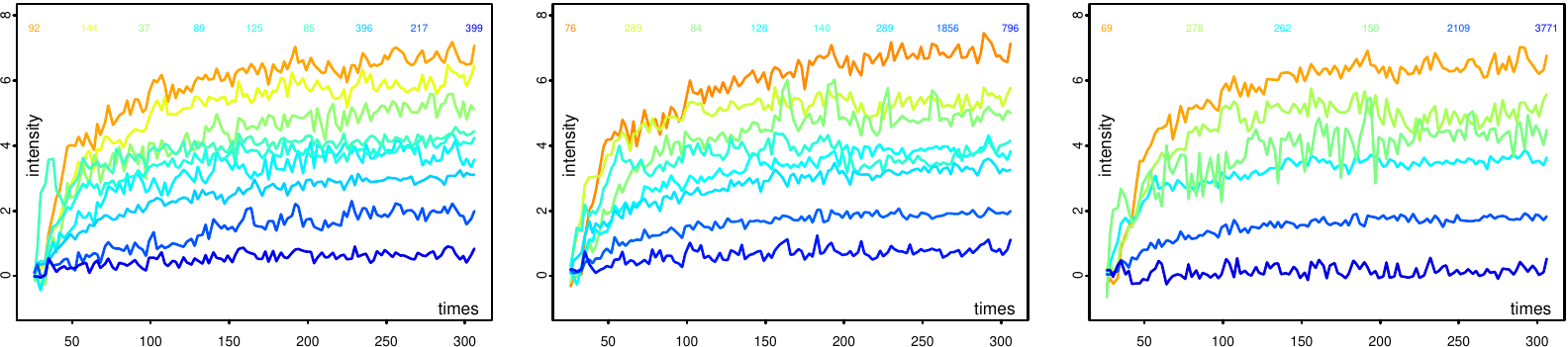}\\\medskip
\rotatebox{90}{\scriptsize \qquad\quad Patient 2}
\includegraphics[height=2.5cm]{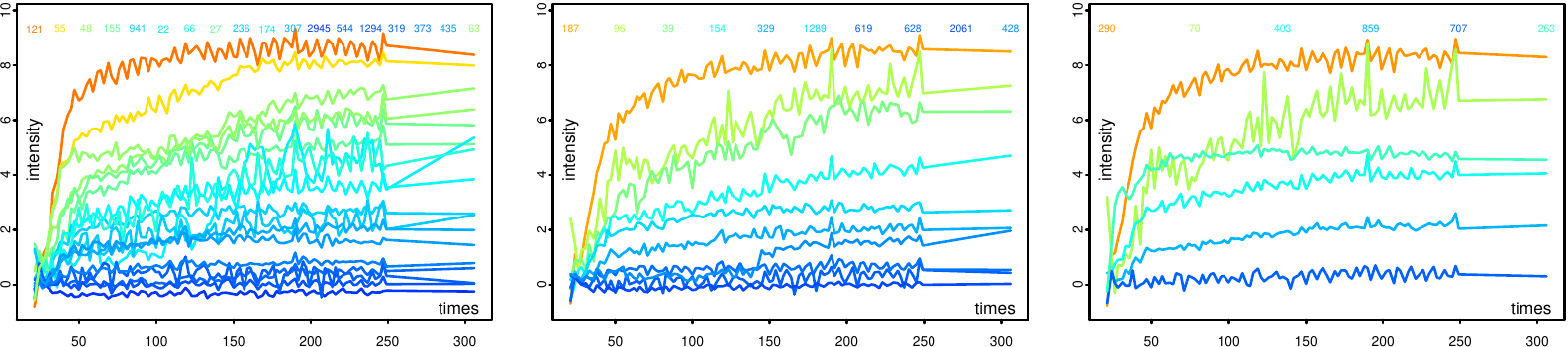}
\caption{Average TCs (after variance stabilization) in the clusters shown inside the manually segmented ROIs (see Fig.~\ref{fig:SegROI}) with $\delta$=2, 3 and 4 from left to right. Size of the corresponding clusters are given at the top of each subfigure with corresponding colors.
}
\label{fig:ROITIC}
\end{figure}

The two real DCE-MR image sequences have been fully segmented into homogeneous regions that are highly consistent with anatomical structures as shown in Figure~\ref{fig:SegRes}. 
The smaller $\delta$ is, the more details in anatomical structures can be observed in image sequence. This is highlighted (see Fig.~\ref{fig:SegROI}) in the ROIs defined by the squares that contain the manually segmented ROIs, shown in black on Figure~\ref{fig:SegRes}. The corresponding estimated TCs inside each cluster within the ROI together with their corresponding size are shown in Figure~\ref{fig:ROITIC}. By segmenting the full DCE sequence, these estimated TCs are obtained by averaging TCs which do not necessarily belong to the ROI but do belong to the same homogeneous cluster. As a benefit, the SNR observed for these estimated TCs is strongly improved, providing a real opportunity for further analysis and comparisons. From these figures, one can clearly understand the advantages of DCE-HiSET. It is indeed providing a piecewise constant representation of the DCE image sequence in functionally homogeneous regions, where $\delta$ controls the size of the pieces and the functional discrepancy between them.

\section{Discussion}

In this paper, we proposed a new method, DCE-HiSET, for segmenting the DCE image sequence into homogeneous regions with respect to the TC observed in each voxel. 
With this approach, the low SNR of DCE image sequence is significantly improved by averaging the TCs of voxels in homogeneous regions. 

Using a dissimilarity measure based on multiple equivalence test, the homogeneity of TC is directly controlled by the equivalence margin, that is the discrepancy tolerance between true but unobservable TCs. The number of regions is automatically determined by this equivalence margin and by the significant level of the testing procedure, the latter having a much smaller influence. At least for synthetic DCE image sequence made up of functionally piecewise homogeneous regions, our algorithm is proven theoretically to be able to retrieve the unknown true partition with high probability as soon as the number of images is large enough.

Consisting of a local and a global clustering step, DCE-HiSET can retrieve the homogeneous regions, highly consistent with real anatomy, regardless of their shape, size and disconnectedness. The total computation complexity is controlled during the local step by the neighborhood structure, and during the global step by the small size of the partitions. 

Through comparison on a synthetic DCE image sequence with a relatively small number of regions, DCE-HiSET outperforms other clustering-based methods with respect to Fowlkes-Mallows indexes and error map. Moreover, our implementation in C++ code wrapped in R is comparable to the best competitors in term of computation time. 

DCE-HiSET can be used for both 2D- and 3D- sequences. However, in 3D, slice thickness and/or distance between slices can be very different from the voxel size in the imaging cross-section. These differences can be taken into account by weighting the $p$-values with respect to the direction of the neighbors.
   
Other extensions of DCE-HiSET include the automatic tuning of parameters $a$ and $\delta$ by optimizing of a simple criterion, which can provide a fully automatic procedure for each DCE image sequence. Used in conjonction with a registration tool \citep{glocker-ar,sotiras-miccai}, DCE-HiSET can provide the right segmentation tool of an iterative registration/segmentation/labelling process during which, along the iterations, both the estimated signals and the registration are learned.

Finally, by only adapting the multiple equivalence test, HiSET can be applied to various types of models where a signal is observed on a spatial field, such as multimodality images, but also such as electrical consumptions observed in a region or a country.

\section*{Acknowledgement}
This work was supported by a CIFRE - ANRT PhD grant in collaboration with Intrasense.

\appendix 
\section{Dyadic decomposition of the time indexes}\label{app1}
We first define an almost regular partition $\mathcal{T}^{K_0}$ of the index set $\{1,\ldots,n\}$ into $2^{K_0}$ sets, with $K_0=\lfloor\log_2 n \rfloor-1$:
\[\mathcal{T}^{K_0} = \left\lbrace T_r, r=1,\ldots,2^{K_0} \right\rbrace \quad \textrm{where}\quad T_r = \left\lbrace j\mid \frac{r-1}{2^{K_0}} < \frac{j}{n} \leqslant \frac{r}{2^{K_0}} \right\rbrace. \]
Then, we build $K_0+1$ partitions of $\{1,\ldots,n\}$, denoted by $\mathcal{T}^K$ and of size $2^K$ for $K=0,\ldots,K_0$, by grouping sets $T_r$ in a pairwise fashion:
\[\mathcal{T}^K := \left\{ T^K_k = \bigcup_{ (k-1)/2^K< r/2^{K_0} \leqslant k/2^K} T_r, \,k=1,\ldots,2^K \right\} .\]
For each $K$, we consider the projection of $D^{XY}$, denoted by $\mathrm{\Pi}_K D^{XY}$, onto the vectors with components constant on each $T_k^K$ of $\mathcal{T}^K$:
\[\mathrm{\Pi}_K D^{XY}:=(\underbrace{m^K_1,\ldots,m^K_1}_{| T^K_1 |},\ldots,\underbrace{m^K_{2^K},\ldots,m^K_{2^K}}_{| T^K_{2^K} |}) ,
\quad m^K_k=\frac{1}{| T^K_k |} \sum_{j\in\mathit{T}_k^K} D^{XY}(t_j).\]
For $K\geq 1$, 
\begin{align*} 
D_K&=(\underbrace{m^K_1-m_1^{K-1},\ldots,m^K_1-m_1^{K-1}}_{| T^K_1 |},\underbrace{m^K_2-m_1^{K-1},\ldots,m^K_2-m_1^{K-1}}_{| T^K_2 |},\ldots,\\
&\qquad\underbrace{m^K_{2^K-1}-m_{2^{K-1}}^{K-1},\ldots,m^K_{2^K-1}-m_{2^{K-1}}^{K-1}}_{| T^K_{2^K-1} |},\underbrace{m^K_{2^K}-m_{2^{K-1}}^{K-1},\ldots,m^K_{2^K}-m_{2^{K-1}}^{K-1}}_{| T^K_{2^K} |}).
\end{align*}
Under Assumption 1, the variance of $m^K_{2k-i}-m^{K-1}_k$, with $k=1,...,2^{K-1}$ and $i=0,1$ is 
\[ (\sigma_{2k-i}^K)^2 = \frac{1}{| T^K_{2k-i} |}+\frac{1}{| T^{K-1}_k |} = \frac{| T^K_{2k-i} |+| T^{K-1}_k |}{| T^K_{2k-i} || T^{K-1}_k |}, \]
and we consider the deterministic diagonal matrix $\Sigma_K^{1/2}$ with diagonal 
\begin{align} \label{eq:Sigma_K}
&\left(\underbrace{\left[\sigma_1^K| T^K_{1} |\right]^{-1},\ldots,\left[\sigma_1^K| T^K_{1} |\right]^{-1}}_{| T^K_1 |},\underbrace{\left[\sigma_2^K| T^K_2 |\right]^{-1},\ldots,\left[\sigma_2^K| T^K_2 |\right]^{-1}}_{| T^K_2 |},\ldots,\right.\\\nonumber
&\left.\qquad\underbrace{\left[\sigma_{2^K-1}^K| T^K_{2^K-1} |\right]^{-1},\ldots,\left[\sigma_{2^K-1}^K| T^K_{2^K-1} |\right]^{-1}}_{| T^K_{2^K-1} |},\underbrace{\left[\sigma_{2^K}^K| T^K_{2^K} |\right]^{-1},\ldots,\left[\sigma_{2^K}| T^K_{2^K} |\right]^{-1}}_{| T^K_{2^K} |}\right)
\end{align}
together with the vector $\bar D_K=\Sigma_K^{1/2} D_K$ called $K$-th rescaled residual after projection. Similarly, for $K=0$, we have 
\[D_0 = (\underbrace{m_1^0,...,m_1^0}_{n})\]
and, as the variance of $m_1^0$ is $1/n$, we define $\bar D_0:=D_0$. 
The $\bar D_{K}$ are independent as the $D_K$ are. Denoting by $\|u\|$ the Euclidian norm of a vector $u$ of $\mathbb R^n$, the normalisation ensures that for $K\geq1$
\begin{align*} 
\| \bar D_K\|^2_n &= \sum_{k=1}^{2^{K-1}} \left( \left[\frac{m^K_{2k-1}-m^{K-1}_k}{\sigma_{2k-1}^K}\right]^2+ \left[\frac{m^K_{2k}-m^{K-1}_k}{\sigma_{2k}^K}\right]^2 \right)  \\
&= \sum_{k=1}^{2^{K-1}} \left(\sqrt{\frac{|T^{K}_{2k-1}||T^{K-1}_{k}|}{|T^{K}_{2k-1}|+|T^{K-1}_{k}|}} (m_{2k-1}^{K} - m_k^{K-1})^2 \right. \\
&\qquad + \left.\sqrt{\frac{|T^{K}_{2k}||T^{K-1}_{k}|}{|T^{K}_{2k}|+|T^{K-1}_{k}|}} (m_{2k}^{K} - m_k^{K-1})^2\right).\end{align*}
taking into account the $2^{K-1}$ linear links, for $k=1,...,2^{K-1}$
\begin{equation} (|T^{K-1}_{2k-1}|+|T^{K-1}_{2k}|)m_k^{K-1} = |T^{K}_{2k-1}| m_{2k-1}^K + |T^{K}_{2k}| m_{2k}^K,\label{eq:link}\end{equation}
 it follows that \[ \|\bar D_K\|^2_n\sim\chi^2(2^{K-1},\| \bar d_K\|^2_n) \] where 
\begin{itemize}
\item $\chi^2(\mu,\lambda)$ is the non-central chi-squared distribution with $\mu$ degrees of freedom and non-centrality parameter $\lambda$,
\item $\bar d_K$ denotes the $K$-th residual after projection of $\bar d=\mathbb E(D)$.
\end{itemize}
For $K=0$, we have moreover $\|\bar D_0\|^2_n\sim\chi^2(1,\| \bar d_0\|^2_n)$.

\section{Proof of Corollary \ref{prop1}} \label{app:proof-prop1} 
Given an observation, the $p$-value is the largest significance level $\alpha$ such that $\mathcal{H}_0$ is accepted. 
Considering rejection region for the IUT of the form $R^\alpha = \bigcap_K R_K^\alpha$, the $p$-value of the IUT  \eqref{IUT} becomes
\begin{align*}
p(X,Y) &= \sup\{\alpha \mid D^{XY} \not\in R^\alpha\} =  \sup \left\lbrace\alpha \mid D^{XY} \not\in \bigcap_K R_K^\alpha\right\rbrace \\
&= \sup \left\lbrace\alpha \mid \bigcup_K \left\lbrace D^{XY} \in \bar R_K^\alpha \right\rbrace \right\rbrace = \sup \left\lbrace\alpha \mid \exists K \text{ with } \underbrace{D^{XY}\in \bar R_K^\alpha}_{p_K(X,Y)\geqslant \alpha}  \right\rbrace \\
&= \sup \left\lbrace\alpha \mid \alpha \leqslant  \max_K(p_K(X,Y)) \right\rbrace = \max_K p_K(X,Y).
\end{align*}

\section{Proof of Theorem \ref{theo1} - Stopping too late}\label{app:proof-theo1}
We want to control $\mathbf{P}_{\mathcal H_0} (p(\bar\ell) \leqslant c_\alpha(\ell))$ the probability of a false merge at iteration $\bar\ell$ with clusters $C_1,\ldots,C_\ell$. This is the probability that 
given $1\leqslant i<j \leqslant \ell$ there exists $K\in \{0,\ldots,K_0\}$ for two clusters $C_i$ and $C_j$ such that they are $\delta$-separated:
\begin{align}
\mathbf{P}_{\mathcal H_0} (p(\bar\ell) \leqslant c_\alpha(\ell))
&= \mathbf{P}_{\mathcal H_0} \left[ \min_{1\leqslant i<j \leqslant \ell} \left\lbrace p(C_i,C_j) \right\rbrace \leqslant c_\alpha(\ell)  \right] \notag\\
&= \mathbf{P}_{\mathcal H_0} \left[ \bigcup_{1\leqslant i<j \leqslant \ell}\left\lbrace p(C_i,C_j) \leqslant c_\alpha(\ell) \right\rbrace \right] \notag\\
&= \mathbf{P}_{\mathcal H_0} \left[ \bigcup_{1\leqslant i<j \leqslant \ell}\left\lbrace \max_K \{p_K(C_i,C_j)\} \leqslant c_\alpha(\ell) \right\rbrace \right] \notag\\
&\leqslant \sum_{1\leqslant i<j \leqslant \ell}\mathbf{P}_{\mathcal H_0} \left[  \max_K \{p_K(C_i,C_j)\} \leqslant c_\alpha(\ell)  \right] \notag\\
&= \sum_{1\leqslant i<j \leqslant \ell}\mathbf{P}_{\mathcal H_0} \left[  \bigcap_K\{p_K(C_i,C_j) \leqslant c_\alpha(\ell)\}  \right]. 
\end{align} 
As the $\bar D^{C_iC_j}_K$ are independent with respect to $K$ by construction (as orthogonal projections of a Gaussian vector, thanks to Cochran's theorem), the $p_K(C_i,C_j)$ are independent. 
Moreover they have the same uniform distribution as $p$-values coming from absolutely continuous distributions. Hence
\begin{align}
\mathbf{P}_{\mathcal H_0} (p(\bar\ell) \leqslant c_\alpha(\ell)) \leqslant  \frac{\ell(\ell-1)}2\prod_K \mathbf{P}_{\mathcal H_0} \left[  p_K(C_1,C_2) \leqslant c_\alpha(\ell)  \right] = \frac{\ell(\ell-1)}2 \left(c_\alpha(\ell)\right)^{K_0+1}.
\end{align} 
Controlling the probability of false merge by $\alpha$ leads to 
\[\frac{\ell(\ell-1)}2 \left(c_\alpha(\ell)\right)^{K_0+1} = \alpha, \]
and we deduce thereby
\[c_\alpha(\ell) = \left(\frac{2\alpha}{\ell(\ell-1)} \right)^\frac{1}{K_0+1}. \]

\section{Proof of Theorem \ref{theo2} - Wrongly binding}\label{app:proof-theo2}
The proof of Theorem \ref{theo2} is a consequence of the Lemma \ref{lem2} which control the probability to bind two $\delta$-separated clusters before non separated ones at a given iteration of the hierarchy. \medskip

\noindent {\it Proof of Lemma \ref{lem2}} \\
We recall that $\bar K:=K-(K>0)$. Assuming that we have at hand a partition $C_1,..., C_\ell$, we consider the sets of couples $(j,j')$ with $1\leq j<j' \leq \ell$ such that $C_j$ and $C_{j'}$
\begin{itemize}
\item have same intensity that we denote $\mathcal S_0$,
\item are $\delta$-separated that we denote $\mathcal D_{\delta}$.
\end{itemize}

We are looking for an upper-bound for the probability to have two $\delta$-separated subsets merging before any two pair of subsets having same intensity, that is
\begin{align*}
P&\left[\min_{\tiny{(j,j')\in\mathcal S_0}} p(C_j,C_{j'}) > \min_{\tiny{(j,j')\in\mathcal D_{\delta}}} p(C_j,C_{j'})\right] \\
&\leq P\left[ (\min_{\tiny{(j,j')\in\mathcal S_0}} p(C_j,C_{j'}) > \min_{\tiny{(j,j')\in\mathcal D_{\delta}}} p(C_j,C_{j'}))  \bigcap (\min_{\tiny{(j,j')\in\mathcal S_0}} p(C_j,C_{j'}) \leq \varepsilon)\right] \\
&\quad+ P\left[(\min_{\tiny{(j,j')\in\mathcal S_0}} p(C_j,C_{j'}) > \min_{\tiny{(j,j')\in\mathcal D_{\delta}}} p(C_j,C_{j'}))  \bigcap (\min_{\tiny{(j,j')\in\mathcal S_0}} p(C_j,C_{j'}) > \varepsilon)\right] \\
&\leq P\left[ (\varepsilon > \min_{\tiny{(j,j')\in\mathcal D_{\delta}}} p(C_j,C_{j'}))  \bigcap (\min_{\tiny{(j,j')\in\mathcal S_0}} p(C_j,C_{j'}) \leq \varepsilon)\right] \\
&\quad+ P\left[\min_{\tiny{(j,j')\in\mathcal S_0}} p(C_j,C_{j'}) > \min_{\tiny{(j,j')\in\mathcal D_{\delta}}} p(C_j,C_{j'})  \Big| \min_{\tiny{(j,j')\in\mathcal S_0}} p(C_j,C_{j'}) > \varepsilon\right]\\
&\quad\times P\left[\min_{\tiny{(j,j')\in\mathcal S_0}} p(C_j,C_{j'}) > \varepsilon\right] \\
&\leq P\left[\varepsilon > \min_{\tiny{(j,j')\in\mathcal D_{\delta}}} p(C_j,C_{j'}) \right] + P\left[\min_{\tiny{(j,j')\in\mathcal S_0}} p(C_j,C_{j'})>\varepsilon\right]\\
&\leq \sum_{\tiny{(j,j')\in\mathcal D_{\delta}}} P\left[\varepsilon >  p(C_j,C_{j'}) \right] +1- P\left[\bigcup_{\tiny{(j,j')\in\mathcal S_0}} p(C_j,C_{j'})\leq\varepsilon\right]\\
&\leq |\mathcal D_{\delta}| P\left[\varepsilon >  p(A,B) \right] +1- P\left[\ p(C,D)\leq\varepsilon\right] \\
&\leq \frac{\ell(\ell-1)}2 P\left[ p(A,B) < \varepsilon\right] +P\left[\ p(C,D)>\varepsilon\right] 
\end{align*}
where $(A,B)$ (respectively $(C,D)$) represents any pair of subsets being $\delta$-separated  (respectively having same intensity) in the partition $C_1,..., C_\ell$.\\

Before to go further, we recall the following concentration inequalities for shifted-$\chi^2$
\begin{prop}[cf Birg\'e]
Let $D$ be a non-central $\chi^2$ variable with $\mu$ degrees of freedom and non-centrality parameter $\lambda\geqslant$ 0, then for all $x>0$
\begin{align}
\mathbf{P}[D\geqslant \mu+\lambda^2+2\sqrt{(\mu+2\lambda^2)x}+2x] &\leqslant e^{-x}, \label{conineq1}\\
\mathbf{P}[D\leqslant \mu+\lambda^2-2\sqrt{(\mu+2\lambda^2)x}] &\leqslant e^{-x}. \label{conineq2}
\end{align}
As a consequence the $\varepsilon$-quantile of a $\chi^2(2^{\bar K},n\delta_K^2)$ satisfies
$$\chi^{-2}_{2^{\bar K},n\delta_K^2}(\varepsilon)\geq 2^{\bar K}+(n\delta_K^2)^2-2\sqrt{(2^{\bar K}+2(n\delta_K^2)^2)\log (1/\varepsilon)}.$$
\end{prop}\bigskip

Let us first focus on the term $P\left[ p(C,D)>\varepsilon\right]$ for $(C,D)=(C_j,C_{j'})$ with $(j,j')\in\mathcal S_0$. By construction and thanks to the independence of the $\bar D_K^{C,D}$
\begin{align*}
P&\left[p(C,D)>\varepsilon\right] = 1-P\left[\max_K P(\chi^2(2^{\bar K},n\delta_K^2)\leq\|\bar D_K^{C,D}\|^2) \leq\varepsilon\right] \\
&=1-\prod_K P\left[ \|\bar D_K^{C,D}\|^2 \leq \chi^{-2}_{2^{\bar K},n\delta_K^2}(\varepsilon) \right] \\
&=1-\prod_K P\left[ \chi^2(2^{\bar K}) \leq \chi^{-2}_{2^{\bar K},n\delta_K^2}(\varepsilon) \right] 
\end{align*}
{
Setting $(n\delta_K^2)^2=a_{K}^{4}2^{\bar K}$ for some $a_{K}$ to be made precise later, we get
\begin{align*}
\lefteqn{P\left[p(C,D)>\varepsilon\right]}\quad\\&\leq 1-\prod_K \left(1-P\left[ \chi^2(2^{\bar K}) > 2^{\bar K}+a_{K}^{4}2^{\bar K}-2\sqrt{(2^{\bar K}+2a_{K}^{4}2^{\bar K})\log\frac1\varepsilon} \right]\right).
\end{align*}
For $K=0,\ldots,K_0$ let us define $y_{K}>0$ as the solution of
\[
2^{\bar K}+2\sqrt{2^{\bar K}y_K}+2y_K=2^{\bar K}+a_{K}^{4}2^{\bar K}-2\sqrt{(2^{\bar K}+2a_{K}^{4}2^{\bar K})\log\frac1\varepsilon},
\]
i.e.
\begin{align}
y_K^2 &= \frac 1 2 \left(\sqrt{2a_{K}^{4}2^{\bar K}+2^{\bar K}-4\sqrt{(2a_{K}^{4}2^{\bar K}+2^{\bar K})\log\frac1\varepsilon}}-2^{\bar K/2} \right)\nonumber\\
&=2^{\bar K/2-1}\left( B_K\sqrt{1-\frac 4 {2^{\bar K/2}B_K}\sqrt{\log\frac1\varepsilon}}-1\right)\label{eq:BK}
\end{align} 
with $B_K^2=4(2a_{K}^{4}+1)$. Then, using the properties of the logarithm, we get
\begin{align}
P\left[p(C,D)>\varepsilon\right]&\leq 1-\prod_K \left(1-e^{-y_K}\right)=1-\exp\left[\sum_{K}\log\left(1-e^{-y_K}\right)\right]\nonumber\\&\leq
1-\exp\left[\log\left(1-\sum_{K}e^{-y_K}\right)\right]=\sum_{K}e^{-y_K}
\nonumber\\&\leq \gamma\frac{e}{e-1}\leq 1.6\, \gamma,\label{eq:CD}
\end{align}
where the first inequality of the last line holds if $y_{K}\ge K-\log\gamma$.
%
Let us now assume that
\begin{equation}
2^{\bar K}B_{K}^{2}\ge(2^{8}/9)\log(1/\varepsilon)\qquad\text{and}\qquad a_{K}^{2}\ge2.
\label{eq-cond1}
\end{equation}
It follows from these conditions that
\[
1-\frac 4 {2^{\bar K/2}B_K}\sqrt{\log\frac1\varepsilon}\ge1/4\qquad\text{and}\qquad
(B_{K}/2)-1=\sqrt{2a_{K}^{4}+1}-1\ge a_{K}^{2},
\]
so that (\ref{eq:BK}) leads to $y_K^2\ge2^{\bar K/2-1}a_{K}^{2}$. Putting all our conditions together, we see that $P\left[p(C,D)>\varepsilon\right]<1.6\gamma$ if
\begin{equation}
(n\delta_K^2)^2=a_{K}^{4}2^{\bar K}\ge\max\left\{2^{\bar K+2};4(K-\log\gamma)^{2};(2^{5}/9)\log(1/\varepsilon)-2^{\bar K-1}\right\}.
\label{eq-ndelta}
\end{equation}
Now, if $A$ and $B$ are two $\delta$-separated subsets then
\begin{align*}
P\left[p(A,B)<\varepsilon\right] &= P\left[ \max(U_0,\ldots,U_{K_0}) < \varepsilon\right]
\end{align*}
where $U_0$, ..., $U_{K_0}$ are $K_0+1$ independent random variables more concentrated on the right of the interval $[0,1]$ than uniforms. We recall that $n/2<2^{K_{0}+1}\leq n$. It follows that
\begin{align*}
P\left[ p(A,B) < \varepsilon\right] &= \prod_{k=0}^{K_0} P\left[ U_k < \varepsilon \right]=\varepsilon^{K_0+1}=2^{-(K_{0}+1)\log_2(1/\varepsilon)}<(n/2)^{\log_2\varepsilon}.
\end{align*}
Hence the overall probability to have two $\delta$-separated subsets merging before any two pair of subsets having same intensity is less than
\[ 
1.6\gamma+\left(\ell^{2}/2\right)(n/2)^{\log_2 \varepsilon}.
\]
At this stage one has the choice of $\gamma$ and $\varepsilon$ but it seems reasonable
to choose the two terms of the previous bound of the same order, i.e.\
$1.6\gamma=(\ell^{2}/2)(n/2)^{\log_2 \varepsilon}$ which leads to the bound $\ell^{2}(n/2)^{\log_2\varepsilon}$. Finally setting $\varepsilon=2^{-\kappa}$
for $\kappa$ large enough, we get the bound $\ell^{2}(n/2)^{-\kappa}$ which can be made small by a convenient choice of $\kappa$. Since $\ell\ge2$, (\ref{eq-ndelta}) holds if 
\[
n\delta_K^2\ge\max\left\{2^{1+\bar K/2};2(K+\kappa\log \frac n 2);1.57\sqrt{\kappa}\right\}.
\]
As $2 \log(n/2)\geq1.57$ for $n\geq 5$ and since $\bar K\leq K_{0}-1\leq\log_2 n-2$, it is enough in this case to have
\[
n\delta_K^2\geq \max\left\{\sqrt{n},2(1+\kappa\log 2) \log_2 \frac n 2\right\}
\]
to ensure the the overall probability to have two $\delta$-separated subsets merging before any two pair of subsets having same intensity is less than $\ell^{2}(n/2)^{-\kappa}$ when $\kappa\ge1$.
} \hfill$\square$\medskip

\noindent {\it Proof of Theorem \ref{theo2}}\\
Along the iterations of the hierarchical clustering, it is enough to bound the probability to wrongly binding at each step using the previous lemma. That is at most $|\mathcal X|$ times. \hfill$\square$\medskip

\section{Algorithm}\label{app:algorithm}
The algorithm consists of two steps: local clustering and global clustering. After distinct definitions of the initial partition $\mathcal P$ and of the neighborhood structure $\mathcal{N}$, two steps share the same main loop to iteratively merge clusters and the same control procedure to stop and to select the number of clusters. 

\paragraph*{Input} $\mathcal{P}$, $\alpha$ and $\delta$.
\paragraph*{Initialization} $\bar p = 0$, $\ell = |\mathcal{P}|$, $\mathcal{N}:=\{\mathcal{V}(C),C\in\mathcal{P}\}$ and $p(X,Y)$ for $X,Y\in \mathcal{P}$.
\paragraph*{Iterations} 
\begin{algorithmic}
\WHILE{$\bar p(\bar\ell) < c_\alpha(\ell)$}
\STATE Find $(s_1,s_2)$ satisfying \eqref{minpair};
\STATE New cluster: $C \leftarrow C_{s_1} \cup C_{s_2}$;
\STATE Update partition: $\mathcal{P}^{\bar\ell+1} \leftarrow \mathcal{P}^{\bar\ell}\setminus \{C_{s_1},C_{s_2}\}\cup\{C\}$;
\STATE New neighbor: $\mathcal V(C):=\mathcal V(C_{s_1}) \cup \mathcal V(C_{s_2}) \setminus  \{C_{s_1},C_{s_2}\}$;
\STATE Compute new dissimilarities as in \eqref{corrdiss} for $C' \in \mathcal{V}(C)$;
\STATE Update $\mathcal{N} \leftarrow \mathcal{N} \setminus \{ \mathcal V(C_{s_1}), \mathcal V(C_{s_2}) \} \cup \mathcal V(C)$;
\STATE $\bar\ell \leftarrow \bar\ell+1$;
\STATE Update $\bar p(\bar\ell)$ as in \eqref{cordisst}.
\ENDWHILE
\STATE Number of clusters: $\ell^{*} \leftarrow N - \bar\ell $.
\end{algorithmic}

\section{HiSET v.s. Classic hierarchical clustering}\label{app:chessboard}

\begin{figure}[h!]
\centering
\includegraphics[width=0.4\textwidth]{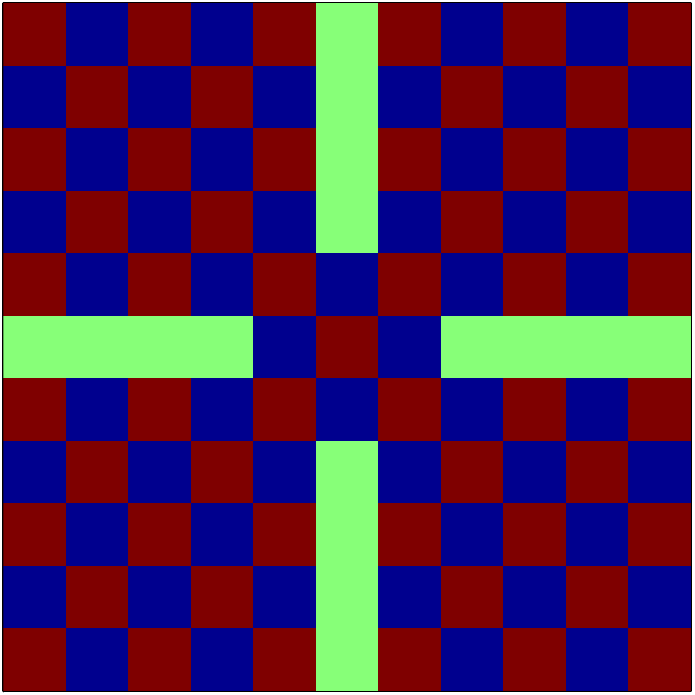} \qquad
\includegraphics[width=0.5\textwidth]{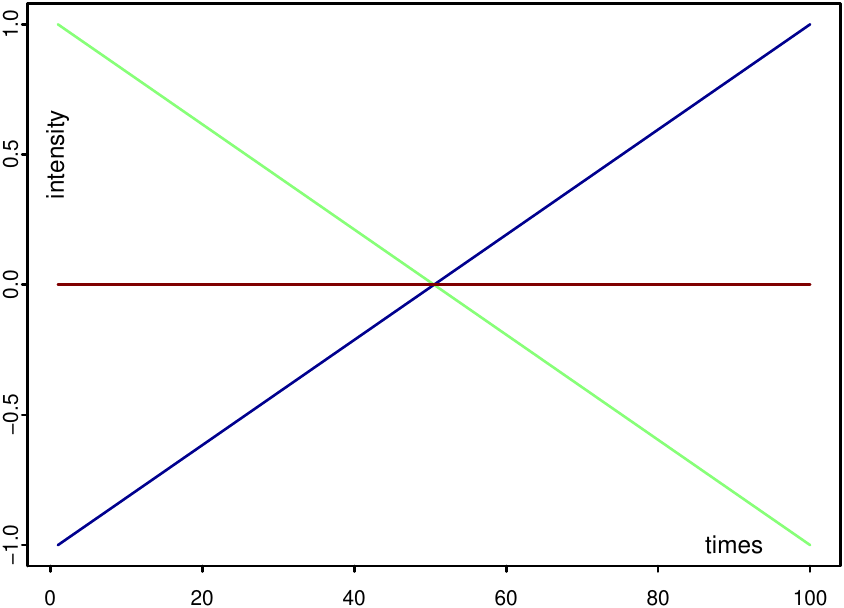}
\caption{Chessboard image sequence: (left) The ground-truth segmentation of $\mathcal X$; (right) The true enhancement curves, $i^x(t)$, associated to the 3 clusters using corresponding colors.}
\label{fig:chessboard}
\end{figure}

In order to demonstrate the strength of HiSET over the classical hierarchical clustering based on the Euclidian distance, we compare both methods on a simulated image sequence consisting of three clusters (see Figure \ref{fig:chessboard}), called latter Chessboard. Standard Gaussian noise has been added to the true enhancement curves to provide the observations. Each image in the sequence is of size $55\times 55$ and the sequence contains 100 images. 

\begin{figure}[h!]
\centering
\includegraphics[width=0.4\textwidth]{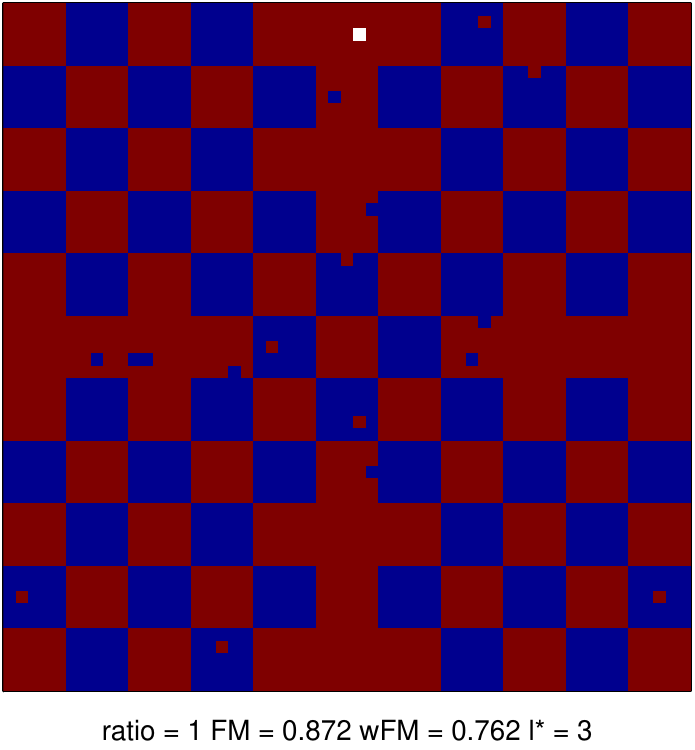} \qquad
\includegraphics[width=0.4\textwidth]{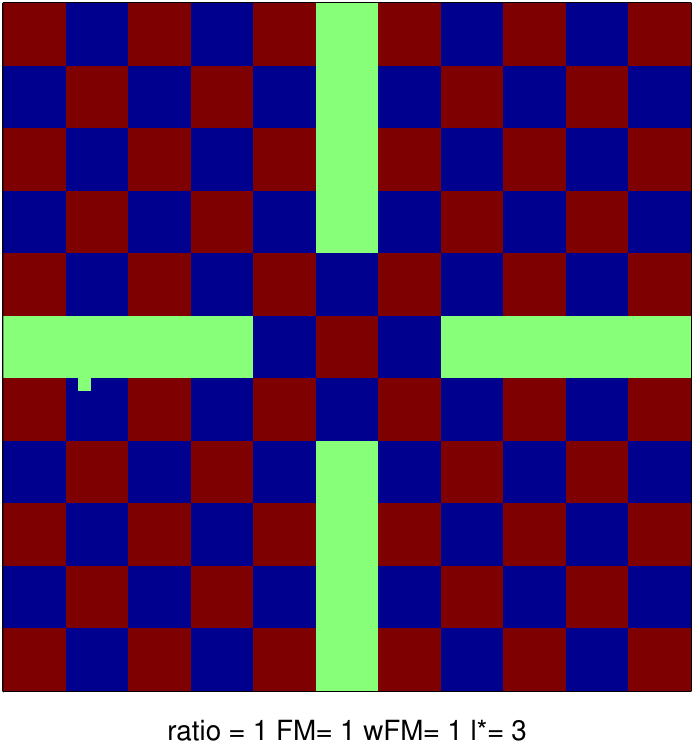}\\[3mm]
\includegraphics[width=0.4\textwidth]{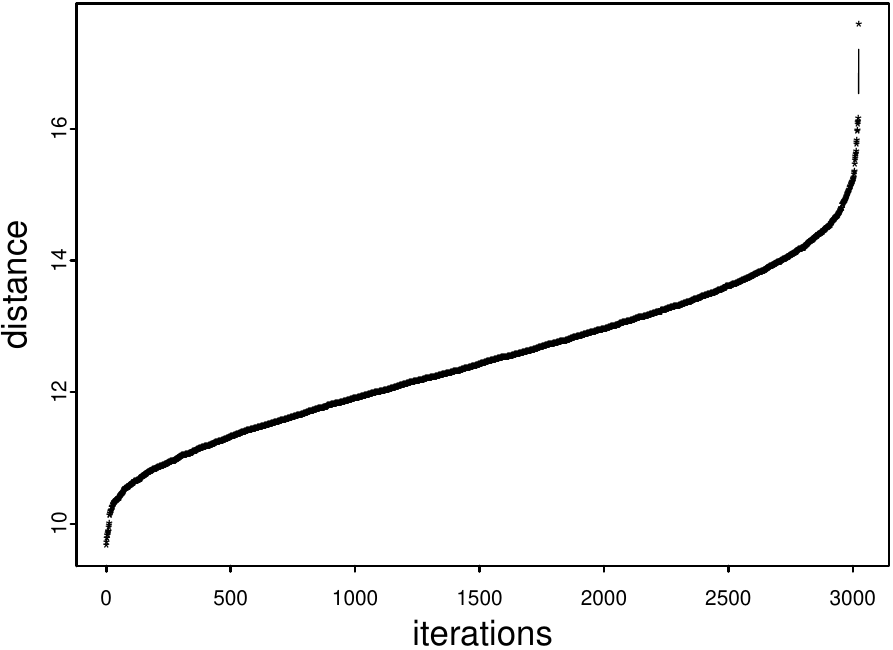} \qquad
\includegraphics[width=0.4\textwidth]{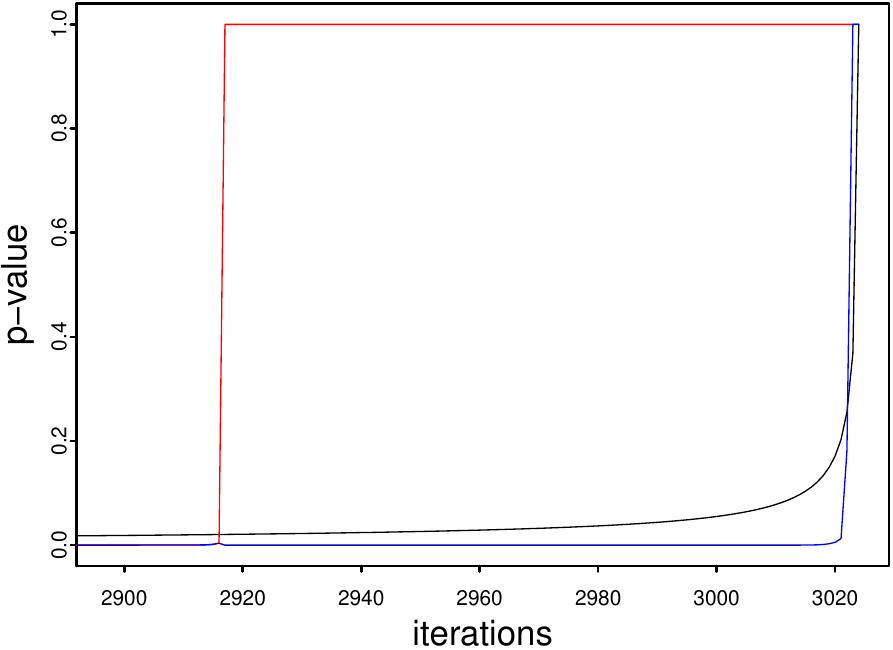}
\caption{Segmentation results for the chessboard image sequence: hierarchical clustering based on the Euclidian distance (left top) and its linkage function (left bottom); HiSET (right top) and the corresponding dissimilarity function together with the control function (right bottom).}
\label{fig:chessboard-seg}
\end{figure}

Figure \ref{fig:chessboard-seg} illustrates the segmentation results.  Given the true number of clusters, the hierarchical clustering based on the Euclidian distance binds red and green clusters and misclassifies several voxels. Indeed the separation distance between the true curves is not large enough with respect to the noise level, providing an illustration of the benefit of using a multiple test. Only, one voxel (in white) is identified as a third cluster. Without any knowledge, HiSET is able to recover all clusters up to a mistake on one voxel.

\section{Notations}\label{app:notations}
\footnotesize\begin{tabular}{cl} 
 $\mathcal X$& the voxel grid corresponding to cross section of the image\\
 $x$, $y$& voxels of the grid\\
 $X$, $Y$& subsets of voxels\\
 $|X|$, $|Y|$& cardinalities\\
 $t_1$, ..., $t_n$& the acquisition times\\
 $\Phi^x(t_j)$& the observed intensity at time $t_j$ and voxel $x$\\
 $\phi^x(t_j)$& the expectation of the intensity at time $t_j$ and voxel $x$\\
 $I^x(t_j)$& the transformed observed intensity at time $t_j$ and voxel $x$\\
$i^x(t_j)$& the expectation of the transformed intensity at time $t_j$ and voxel $x$\\
$\eta^x_j$& an additive standard Gaussian noise\\
$\bar I^X$& empirical mean of the $I^x$ over $X$\\
$\bar i^x$& expectation of the empirical mean ofver $X$\\
$\rho(X,Y)$& scaling factor $|X|^{-1}+|Y|^{-1}$\\
$D^{XY}$& normalized difference of $\bar I^X$ and $\bar I^Y$\\
$d^{XY}$& expectation of normalized difference of $\bar i^X$ and $\bar i^Y$\\
$K$& a time index partition reference\\
$K_0$& the reference of the finest time index partition\\
$\mathrm{\Pi}_K D^{XY}$& projection over the $K$-th partitition\\
$D_{K}^{XY}$& $\Pi_K D^{XY}-\Pi_{K-1} D^{XY}$\\
$\Sigma_K$& covariance matrix of $D_{K}^{XY}$\\
$\bar D_K$& scaled version of $D_{K}^{XY}$\\
$\bar d_K$& expectation of $\bar D_K$\\
$\|u\|_n$& Euclidian norm of $u$ in $\mathbb  R^n$\\
$C_1$, ..., $C_\ell$& some clusters\\
$\mathcal N(\mu,\sigma^2)$& Gaussian distribution with mean $\mu$ and variance $\sigma^2$\\
$\mathcal N(d,\Sigma)$& multivariate Gaussian distribution with mean $d$ and covariance matrix $\Sigma$\\
$\chi^2(d,\lambda)$& shifted chi-squared distribution\\
$\mathcal H_0^K$& the null hypothesis of the equivalence test over the $K$-th partition\\
$\mathcal H_1^K$& the alternative of the equivalence test over the $K$-th partition\\
$p_K(X,Y)$& the $p$-value over the $K$-th partition when testing $\bar i^X(.)=\bar i^Y(.)$\\
$p(X,Y)$& the $p$-value of the multiple equivalence test of $\bar i^X(.)=\bar i^Y(.)$\\
$\bar\ell$& $|\mathcal X|-\ell$\\
$p(\bar\ell)$& the minimum dissimilarity at iteration $\bar\ell$\\
$c_\alpha(\ell)$& the control function when the clustering is made of $\ell$ clusters (iteration $\bar\ell$)\\
$\mathcal P^{loc}$& the final partition after the local step\\
$\ell^{loc}$& the size of $\mathcal P^{loc}$\\
$\mathcal P^{*}$& the final partition after the global step\\
$\ell^{*}$& the size of $\mathcal P^{*}$\\
$\ell_0$& the size of the true underlying partition\\
\end{tabular}

\normalsize

\newpage

\bibliography{mybibfile}

\end{document}